\documentclass[preprint,12pt]{elsarticle}

\usepackage{amssymb}
\usepackage{mathtools}
\usepackage{subfig}
\usepackage{caption}
\usepackage{xcolor}
\usepackage{tcolorbox}
\usepackage{amsthm}
\usepackage{amsmath}
\usepackage{natbib}
\usepackage{svg}
\usepackage{subcaption}
\usepackage{multirow}

\usepackage{setspace}
\usepackage[a4paper, left=2cm, right=2cm, top=2.5cm, bottom=2.5cm]{geometry}

\journal{Journal of Colloid and Interface Science}

\begin{document}
	
	\begin{frontmatter}
		
\author[inst1]{Afsal Chakkam Palliyalil} 

\author[inst1]{Gaurav Tomar\corref{cor1}}
\ead{gtom@iisc.ac.in}

\author[inst1]{Susmita Dash\corref{cor1}}
\ead{susmitadash@iisc.ac.in}

\cortext[cor1]{Corresponding author}

\affiliation[inst1]{organization={Department of Mechanical Engineering}, addressline={Indian Institute of Science}, city={Bangalore}, country={India}}
		
\title{Universality of Bubble Coalescence in Electrolytic Media}
		
\begin{abstract}
Bubble coalescence phenomenon in electrolytic media finds applications in technologies from mineral flotation to electrochemical energy conversion. However, the underlying governing physics still remains unresolved, with longstanding disagreement over the extent to which Marangoni stresses affect the coalescence time by modulating the interfacial mobility. Here, we show that the thin film morphology governs drainage more strongly than the interfacial boundary conditions. We demonstrate experimentally that thin film drainage during bubble coalescence proceeds through three distinct regimes. An initial visco-capillary stage that exhibits a power-law thinning, followed by an exponential decrease in film thickness with time induced by rim stabilisation. The final regime is governed by disjoining pressure and is marked by an exponential relaxation of the film to the equilibrium thickness. We show that, irrespective of the electrolyte type and concentration, film evolution exhibits universal behavior by collapsing onto a single curve when rescaled with the characteristic film thickness and time scale, demonstrating that electrolyte effects act only to renormalize timescales rather than alter the underlying dynamics.  
\end{abstract}
            
\end{frontmatter}
	
\noindent
\section*{Introduction}
Bubble coalescence is ubiquitous in natural and industrial processes ranging from air–sea gas exchange, embolism in plant xylem, humans and marine animals to mineral flotation, foam stability, water treatment, and microfluidic operations \citep{deike2025universal,grossiord2020plant,AHMED1985195,PhysicsFoamsWeaire,Jia2023}. In all of these systems, the stability of the thin liquid film trapped between two approaching interfaces controls coalescence and therefore the rates of mass, heat, and charge transport in multiphase environments \citep{Kantarci2005,JalalInanlu2024,Zhang2024}. The drainage of the intervening liquid film is strongly influenced by the transport properties of the liquid and interface deformation. As a bubble approaches another interface, the hydrodynamic pressure generated within the draining film can exceed the bubble Laplace pressure, producing a characteristic dimple morphology consisting of a central thick region surrounded by an annular rim where the film is thinnest \citep{hartland1977model,tsekov1994dimple}. 

While pure liquids are generally incapable of sustaining stable clusters of bubbles, saline environments, exemplified by breaking ocean waves, exhibit persistent frothing and delayed bubble coalescence \citep{deane2002scale,katsir2015bubble}. Although surface-active organic matter contributes to foam stability through adsorption at bubble interfaces, comparable concentrations of dissolved organic matter are also found in freshwater systems that do not exhibit persistent foaming \citep{schilling2011foam,mostofa2012dissolved}. These observations indicate a distinct role of dissolved electrolytes in modifying bubble coalescence dynamics.

Bubble coalescence inhibition by dissolved electrolytes remains one of the long-standing unresolved problems in interfacial science. Since the first observations of electrolyte-induced inhibition of coalescence nearly a century ago \citep{Schnurmann1929}, a wide range of mechanisms have been proposed, yet no consensus has emerged regarding the physical origin of the phenomenon. Because bubble coalescence requires drainage and rupture of a thin liquid film, early interpretations adopted the framework of colloidal film stability in which electric double-layer (EDL) repulsion stabilizes the film \citep{prince1990transition}. 
However, increasing electrolyte concentration enhances screening of electrostatic interactions, reducing the range of EDL repulsion and decreasing the equilibrium film thickness. Delayed coalescence, however, is frequently observed under conditions where electrostatic interactions are strongly screened \citep{craig1993effect}, suggesting that classical EDL interactions alone are insufficient to explain the  sensitivity of coalescence dynamics to electrolyte composition.

An alternative explanation was proposed by Marrucci \citep{marrucci1967coalescence}, who argued that redistribution of ions during film drainage generates surface tension gradients that oppose liquid motion through Gibbs–Marangoni stresses. This framework established the basis for many subsequent interpretations of electrolyte-induced coalescence inhibition. The importance of ion-specific effects was later highlighted by Craig et al. \citep{craig1993effect,craig2004bubble}, who demonstrated that coalescence inhibition depends not only on the ionic strength but also on the specific combination of ions present in the solution. Their classification of ions into surface-depleting ($\alpha$) and surface-enhancing ($\beta$) groups revealed empirical combining rules in which $\alpha$–$\alpha$ and $\beta$–$\beta$ electrolytes inhibit coalescence whereas $\alpha$–$\beta$ and $\beta$-$\alpha$ combinations exhibit little or no effect. These observations pointed toward a central role of ion-specific interfacial partitioning that could not be readily reconciled within the conventional DLVO theory. Craig further noted that the Gibbs–Marangoni mechanism contains an inherent feedback: the Marangoni stresses reduce the interfacial mobility thereby also suppressing the generation of concentration gradient required to sustain them \citep{craig2004bubble}. More recently, it has been shown that ion-specific surface potentials arising from unequal interfacial ion partitioning may provide a route toward understanding the observed $\alpha$–$\beta$ behaviour \citep{duignan2021surface}.
Collectively, these studies have progressively shifted attention away from equilibrium interaction forces toward ion-specific interfacial organization and dynamic transport processes \cite{li2025liquid}.

Recent experimental advances have provided  greater insights into the dynamics of thin-film drainage. Interferometric studies have demonstrated that coalescence time depends strongly on drainage kinetics, bubble size and approach velocity rather than solely on equilibrium interaction forces \citep{christenson2008electrolytes,horn2011coalescence}. It has been shown that  pristine air-water interfaces remain highly mobile during drainage and that even trace levels of contamination can substantially modify film evolution and coalescence behaviour \citep{manica2018dynamic,carnie2019mobile}. These studies have established that bubble coalescence is governed by a strong coupling between interfacial mobility and hydrodynamic drainage. High-speed interferometric measurements subsequently revealed that films of electrolyte solutions initially drain in a manner nearly identical to pure water before undergoing a sudden slowdown when the thickness reaches $\sim 30$--$50$ nm \citep{li2025liquid, liu2023nanoscale}. These observations motivated nanoscale transport-based models in which delayed coalescence is due to Gibbs-Marangoni pressure affecting the coupled bulk ion transport and interfacial surface excess. Although these models capture the overall drainage behavior in experiments, electrolyte-dependent variations in drainage kinetics also occur at film thicknesses substantially larger than the confinement scales predicted to trigger nanoscale transport effects \citep{christenson2008electrolytes,horn2011coalescence}. Consequently, the relationship between ion-specific interfacial transport, electrostatic interactions, film drainage and film morphology remains incompletely understood.

Several theoretical and experimental studies suggest that electrolyte-induced stabilization arises from a complex interplay between interfacial ion transport, Gibbs–Marangoni stresses and ion-dependent modifications of interfacial electrostatics \citep{firouzi2017gibbs,chatzigiannakis2024perspective,sharma2025foam}.
Surface-enhancing and surface-depleting ions modify the interfacial charge distribution in fundamentally different ways and may therefore alter both interfacial transport and electrical double-layer structure. However, existing descriptions generally treat electrostatic interactions through an average film thickness or a characteristic disjoining pressure. Whether electrolyte-dependent EDL effects act primarily through the minimum film thickness at the rim, rather than through the central film region, remains unexplored. Resolving this question is essential for connecting ion-specific interfacial physics with the hydrodynamic evolution of deformable thin films.

Here, we use colour interferometry to track the spatiotemporal evolution of the liquid film entrapped between a millimetric air bubble and a flat glass substrate in monovalent (NaCl) and divalent (Na$_2$SO$_4$) electrolytes with approximately 20 nm thickness resolution. We show that increasing electrolyte concentration and ion valency systematically reduce the rim thickness while simultaneously slowing film drainage. Analysis of the evolving film morphology reveals three distinct drainage regimes spanning visco-capillary deformation, rim-controlled relaxation and nanometric film stabilization. By identifying the characteristic scales governing each regime, we derive scaling laws that quantitatively describe the drainage dynamics across electrolyte type, concentration and film thickness. The resulting collapse of our measurements and previously reported experimental data onto a universal master curve provides a unified framework linking electrolyte-dependent electrostatic interactions, rim evolution and thin-film drainage dynamics.

\section*{Results and discussion} \label{sec_Results}
\subsection*{Spatio-temporal evolution of film drainage}

	\begin{figure}
		\centering		
		\includegraphics[width=\linewidth]{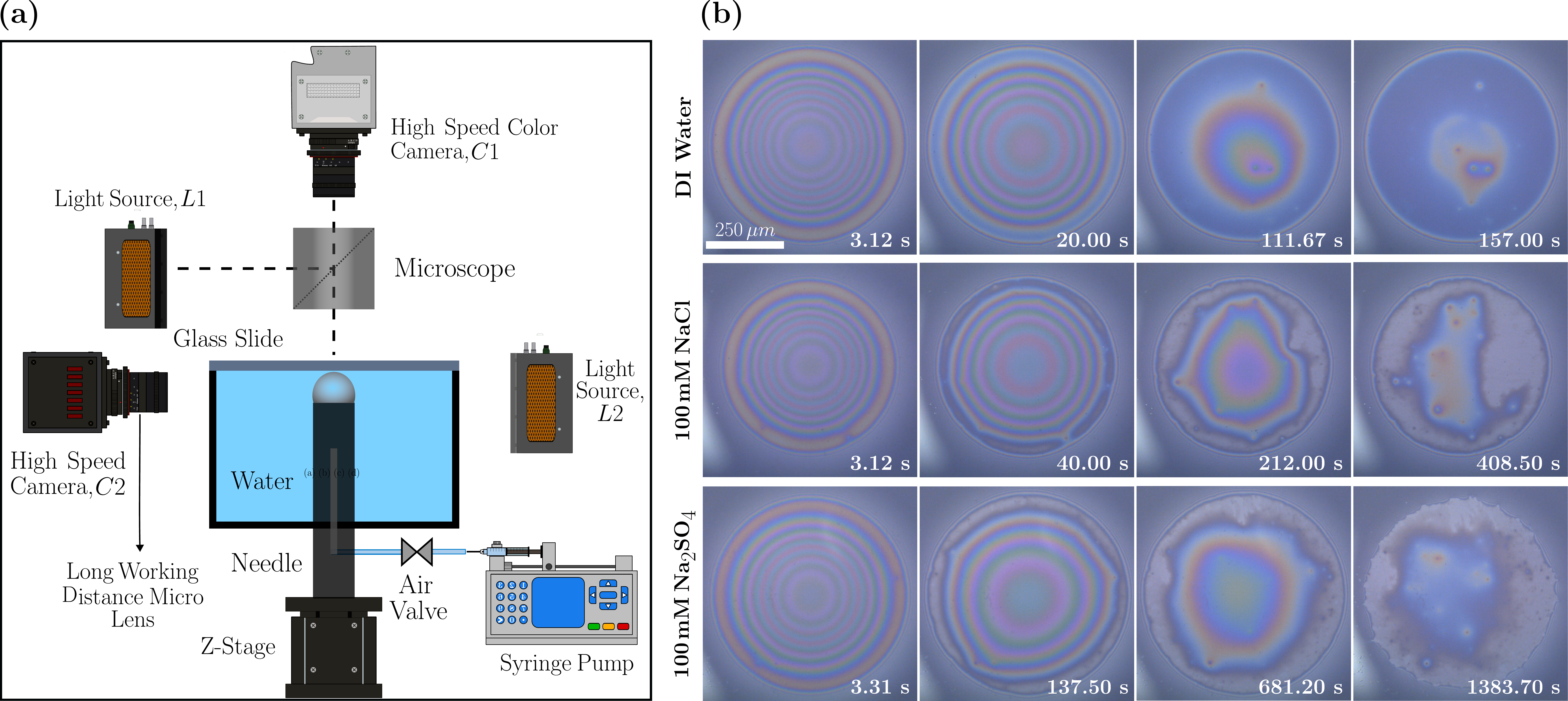}
		\caption{Experimental configuration and evolution of interference fringes during thin-film drainage between a rising bubble and a glass substrate. (a) Schematic of the optical interferometry setup used to capture the spatio-temporal evolution of the entrapped liquid film. (b) Time-resolved color interferograms for three representative systems: deionized (DI) water (row~1), NaCl (100~mM, row~2), and Na$_2$SO$_4$ (100~mM, row~3). Each row shows successive stages of film evolution with timestamps indicating the elapsed time after the onset of the bubble approach.} \label{fig_ExpSetup_Fringes}
	\end{figure}
We employ high-speed color interferometry to analyse the evolution of the thin liquid film entrapped between an approaching air bubble and a glass substrate, as illustrated in Figure~\ref{fig_ExpSetup_Fringes}(a) (see Methods for detailed experimental procedures). The interference patterns provide direct visual access to the spatio-temporal evolution of the film thickness and its dependence on electrolyte conditions. Figure~\ref{fig_ExpSetup_Fringes}(b) presents representative interferograms for three systems: deionized (DI) water, 100~mM NaCl, and 100~mM Na$_2$SO$_4$. Within each row, the images correspond to successive time instants of film drainage; $t = 0$ corresponds to the instant at which the bubble begins to approach the glass substrate from an initial separation of $601 \pm 5 \; \mu$m. While the overall drainage dynamics are significantly slowed in the presence of electrolyte as compared to DI water for both the electrolytes, similar fringe morphologies are observed at different times, revealing a common sequence of dimple formation, rim thinning, and eventual nanometric film stabilization. 

During the early stage of drainage, the interference fringes appear as nearly perfect concentric rings, indicating an axisymmetric dimple and spatially uniform thinning of the liquid film. This behavior is consistent across DI water and all electrolyte solutions (NaCl and Na$_2$SO$_4$ at 1, 10, and 100~mM; see Supplementary Figures ~S3 and S4). As the film thins further, the fringe spacing becomes angularly non-uniform, and the number of visible fringes progressively decreases, while each fringe becomes radially broader. Simultaneously, the outermost fringe, serving as the effective optical signature of the dimple rim, undergoes pronounced evolution during drainage. While its outer boundary remains nearly stationary, its inner edge progressively shifts toward the center of the interferogram, signifying a continuous increase in the dimple shoulder width, w. For instance, for DI water, the width of the outermost fringe increases from approximately 44~$\mu$m at 3.12~s to approximately 108~$\mu$m at 111.67~s. Together, these changes reflect a flattening of the dimple profile and a reduction in curvature, consistent with the overall geometric relaxation of the draining film observed across all concentrations.

The radial spreading of the thin liquid film confined between the bubble and the glass substrate continues till the $z$-stage stage motion is stopped at $t = \Delta h / V_b = 2.15 \pm 0.3$~s (details in Methods section). Notably, the lateral extent of the interferogram is nearly identical $\sim590 \pm 30~\mu$m across all the solutions. This indicates that the ultimate radial footprint of the entrapped film (see Figure~\ref{fig_ExpSetup_Fringes}(b) and Figures~S4 and S5, at $t = 200$~s) is governed primarily by the imposed approach displacement $\Delta h$ and velocity, which sets the geometric extent of bubble deformation.

\subsection*{Ion-specific retardation of drainage kinetics}
      
\begin{figure}
\centering
\includegraphics[width=1\linewidth]{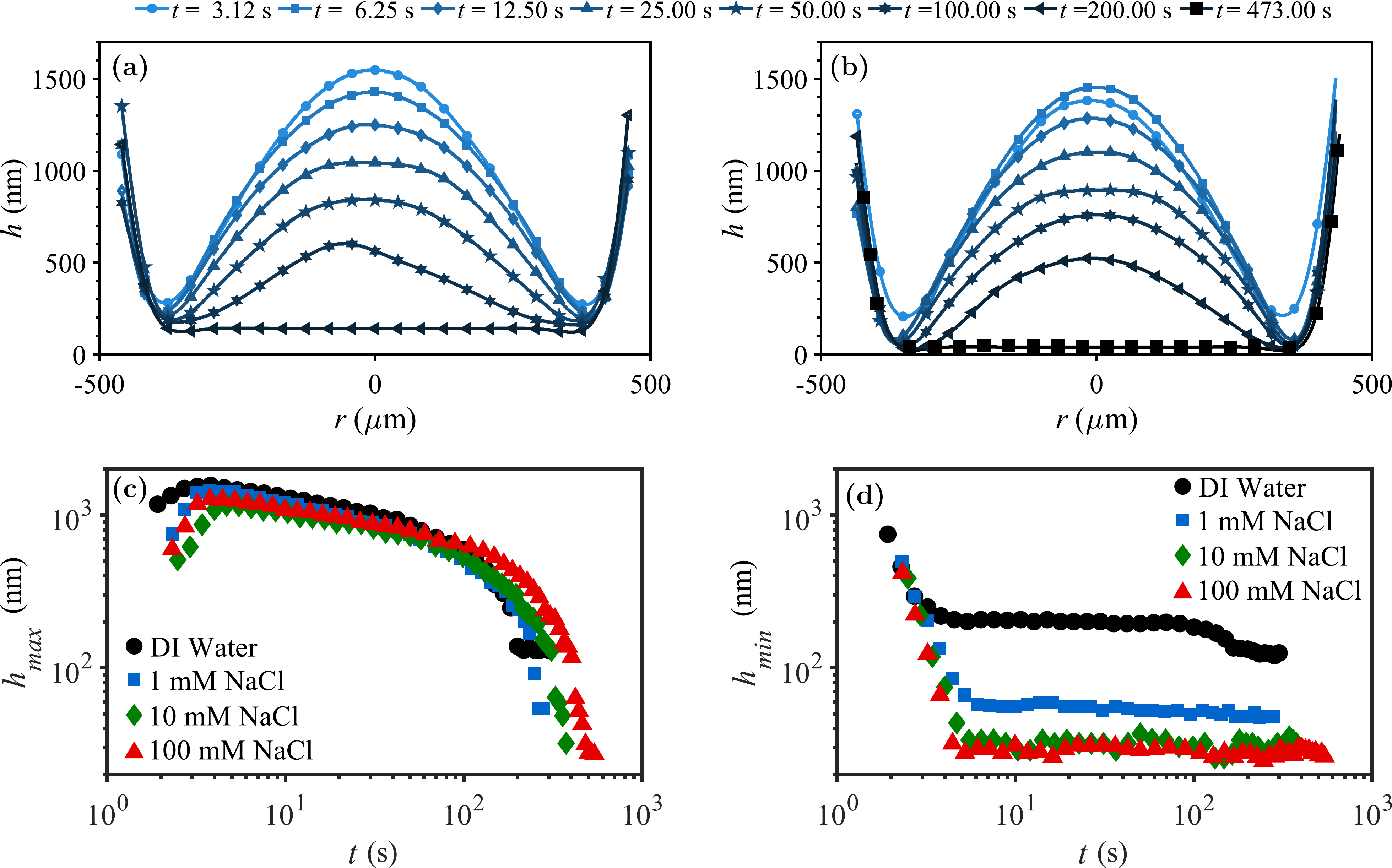}
\caption{Electrolyte-dependent drainage dynamics of the thin liquid film formed between a bubble ($R_b = 1.39$~mm) and a flat glass substrate. Panels (a) and (b) show the spatio-temporal evolution of the film thickness profiles, $h(r,t)$, for (a) deionized (DI) water and (b) NaCl (100~mM). The profiles at successive time instants (see legend) capture the progressive dimple formation and rim thinning, with markedly slower drainage and enhanced deformation at higher ionic strength. Panels (c) and (d) present the temporal evolution of key geometric metrics across DI water and NaCl solutions (1, 10, and 100~mM); (c) maximum dimple height, $h_{\max}$, and (d) minimum rim height, $h_{\min}$.}
\label{fig_Collated_NaCl_hvsr_1_100mM_hmax_hmin}
\end{figure}

To isolate the influence of ionic strength on the dimple geometry and its drainage, we first evaluate the drainage profiles of monovalent NaCl solutions. Figure~\ref{fig_Collated_NaCl_hvsr_1_100mM_hmax_hmin}(a,b) shows the spatio-temporal evolution of the film thickness, $h(r,t)$, as a function of the radial position $r$, for DI water and NaCl solutions of concentration 100~mM; film evolution for 1 and 10 mM concentration is provided in Supplementary Figure S6. At early times, all the solutions exhibit nearly identical drainage behavior independent of electrolyte concentration. As drainage proceeds, the profiles exhibit a progressive reduction in the minimum film thickness at the annular rim, accompanied by a gradual flattening of the central dimple. Eventually, the maximum dimple thickness approaches the rim thickness, leading to an almost spatially uniform film profile, albeit at different time instants ($t_{stable}$) as illustrated in Figure~\ref{fig_Collated_NaCl_hvsr_1_100mM_hmax_hmin}(a,b).
This geometric flattening of the film signals the onset of stabilization; any subsequent thinning proceeds uniformly throughout the film without pronounced radial variations. The time required for the film to attain a quasi-uniform configuration increases monotonically with NaCl concentration. The stabilization time rises from approximately $198~\mathrm{s}$ for DI water to approximately $473~\mathrm{s}$ for the $100~\mathrm{mM}$ NaCl solution (Table~\ref{tab_stabilised_para_combined}).

Figures~\ref{fig_Collated_NaCl_hvsr_1_100mM_hmax_hmin}(c,d) show the temporal evolution of the maximum central thickness (dimple height), $h_{\max}$, and the minimum rim height, $h_{\min}$ for DI water and different concentrations of NaCl solution. The corresponding evolution of the dimple height, $h_{\mathrm{dimp}}(t)$, and rim radius, $r_{\mathrm{rim}}(t)$, is presented in Supplementary Figure~S6. Although the $z$-stage motion ceases at $t=2.15$~s, the central thickness continues to increase for an additional $\sim0.5$--$0.75$~s. For the baseline case of DI water, $h_{\max}$ reaches its maximum value ($\sim1600$~nm) at $\sim 3$--$4$~s, followed by thinning until approximately $\sim 200$~s to reach the equilibrium film thickness (Figure~\ref{fig_Collated_NaCl_hvsr_1_100mM_hmax_hmin}(c)). We note that in the log-log graph, $h_{max}$ decreases linearly initially till $\sim 66.6$s and deviates thereafter. 
The corresponding film thickness at the rim, $h_{\min}$, rapidly decreases initially to about $\sim200$~nm within the initial $5$~s and subsequently decreases at a significantly slower rate to reach an equilibrium film thickness of $\sim 129$ nm at $\sim 200$ s, where $h_{min} \sim h_{max}$(Figure~\ref{fig_Collated_NaCl_hvsr_1_100mM_hmax_hmin}(d) and Figure~S7(b)).

However, increasing the NaCl concentration reveals systematic deviations from this baseline behavior. While all NaCl solutions (1, 10, and 100~mM) initially exhibit a qualitatively similar evolution of $h_{max}$, the key drainage transitions, including the termination of the initial linear thinning regime (in the log-log plot) and the eventual stabilization of the film, occur at progressively delayed times with increasing electrolyte concentration.  The maximum film thickness, $h_{\max}$, attains peak values of $1557$, $1155$, and $1264$~nm at roughly the same time ($3$--$4$~s) for $1, 10$ and $100$ mM NaCl solutions, respectively. Following these peak values, $h_{\max}$ decreases rapidly until approximately 47~s, 58~s, and 80~s to $\sim$834, 673, 675 nm for the three concentrations, respectively (Figure~\ref{fig_Collated_NaCl_hvsr_1_100mM_hmax_hmin}(c)), marking the end of the first drainage regime. Beyond these times, $h_{\max}$ exhibits a slow decay similar to DI water, although with progressively lower equilibrium thicknesses with increasing concentration as listed in Table~\ref{tab_stabilised_para_combined}. The evolution of $h_{\min}$, follows a trend similar to DI water, although with significantly lower values (see Figure \ref{fig_Collated_NaCl_hvsr_1_100mM_hmax_hmin}(d)).  Subsequently, the thin film equilibrates to a nearly flat film (see Figure \ref{fig_Collated_NaCl_hvsr_1_100mM_hmax_hmin}(b) for 100mM NaCl solution). 

\begin{table}
\caption{Debye length $\lambda_D$, stabilization time $t_{\mathrm{stable}}$, and stabilized film thickness $h_{\mathrm{stable}}$ for DI water, NaCl, and Na$_2$SO$_4$ solutions at different concentrations.}
\centering
\begin{tabular}{l l c c c}
\hline\hline
Electrolyte & Concentration & $\lambda_D$ (nm) & $t_{\mathrm{stable}}$ (s) & $h_{\mathrm{stable}}$ (nm) \\[0.5ex]
\hline
DI Water     & --      & --    & 198  & 129 \\
\hline
NaCl         & 1 mM    & 9.612 & 251  & 54  \\
NaCl         & 10 mM   & 3.040 & 323  & 32  \\
NaCl         & 100 mM  & 0.961 & 473  & 27  \\
\hline
Na$_2$SO$_4$ & 1 mM    & 5.550 & 401  & 40  \\
Na$_2$SO$_4$ & 10 mM   & 1.754 & 660  & 26  \\
Na$_2$SO$_4$ & 100 mM  & 0.555 & 1680 & 21  \\
\hline
\end{tabular}
\label{tab_stabilised_para_combined}
\end{table}

The final stable film thickness, $h_{stable}$, reduces progressively from $\sim 54$~to~$\sim 27$ nm as the electrolyte concentration increases from $1$ to $100$ mM. The electrostatic interactions arising from the electrical double layer (EDL) play a significant role in determining $h_{stable}$ \cite{israelachvili1992intermolecular, pushkarova2005surface}. As electrolyte concentration increases, the Debye length, $\lambda_D$ decreases ($\lambda_D \propto 1/\sqrt{I}$, where $I$ is the ionic strength), leading to stronger screening of surface charges \cite{derjaguin1974structural}. This reduces the repulsive disjoining pressure at the film shoulder. The consequent reduction in the annular drainage area at the rim confines the outward radial flow and effectively traps a larger volume of fluid beneath the bubble, thereby significantly retarding drainage as evident from Figure~\ref{fig_Collated_NaCl_hvsr_1_100mM_hmax_hmin}(c). Thus, we note that the rim region, that is, both $h_{min}$ and $r_{rim}$  stabilizes while $h_{max}$ continues to evolve before the film reaches an eventual thickness of $h_{stable}$.

The influence of NaCl concentration is reflected in the temporal evolution of the dimple height. At $t=100$~s, the measured values are, $
h_{dimp,\mathrm{DI}} = 389~\mathrm{nm},
h_{dimp,1\mathrm{mM}} = 466~\mathrm{nm}$, $
h_{dimp,10\mathrm{mM}} = 503~\mathrm{nm}$ and $
h_{dimp,100\mathrm{mM}} = 593~\mathrm{nm}.
$
Thus, the dimple in the $100$~mM NaCl solution is nearly $66\%$ larger than in DI water, demonstrating  significant retardation of film drainage in the presence of electrolytes.

To elucidate the role of ion valency, analogous experiments were performed with Na$_2$SO$_4$ solutions (Supplementary Figures~S5 and S7). While the early-time drainage dynamics remain indistinguishable from DI water and NaCl, consistent with similar interfacial boundary conditions, pronounced deviations emerge at later times. In contrast to NaCl, Na$_2$SO$_4$ solutions exhibit substantially prolonged drainage, sharper and more confined rim structures, and significantly delayed film stabilization, with the effect strongly amplified at higher concentrations. For instance, for 100 mM concentration, $t_{stable}$ increases ~3.6 times for Na$_2$SO$_4$ compared to NaCl solution (see Table \ref{tab_stabilised_para_combined}). We note that for Na$_2$SO$_4$, the shorter Debye length is significantly more screened due to higher SO$_4^{-2}$ ion valency. 

These observations confirm that although the overall drainage time is significantly different for different salts and their concentrations, the dynamics remain qualitatively similar. Increasing NaCl/Na$_2$SO$_4$ concentration prolongs the persistence of the central dimple, accelerates thinning of the rim region, and modifies the lateral drainage profile. Consequently, increasing electrolyte concentration progressively alters the spatio-temporal evolution of the thin film.

 The delay in coalescence (slower drainage of thin intervening films) in electrolyte solutions has been attributed to the Gibbs Marangoni stress arising from non-uniform spatial distribution of ions at the interface \cite{li2025liquid, liu2023nanoscale, liu2019coalescence}.  The local surface tension increases compared to that of the bulk in confined liquid films of thickness $h$ is given as \cite{marrucci1969theory},
\begin{equation}
\Delta \sigma = \frac{c}{N_A k_B T\, h} \left(\frac{d\sigma}{dc}\right)^2,
\label{Eqn_STMarrucci}
\end{equation}
where $c$ is the bulk electrolyte concentration, $k_B T$ is the thermal energy, and $d\sigma/dc$ is the variation in surface tension with concentration. Equation~\ref{Eqn_STMarrucci} shows that $\Delta\sigma \propto h^{-1}$, implying that surface tension increases with decrease in film thickness.  Using $d\sigma/dc \approx 1.787~\mathrm{mN\,m^{-1}\,M^{-1}}$ \cite{pegram2006partitioning} for 100 mM NaCl, the predicted surface-tension increase at the dimple center ($h\approx500$~nm) is negligible, $\Delta \sigma \approx 0.0002 ~ \mathrm{mN\,m^{-1}}$, whereas at the thinned rim ($h\approx30$~nm), the local surface tension increases by $\Delta \sigma \approx 0.0043 ~ \mathrm{mN\,m^{-1}}$, an order-of-magnitude increase that generates significant Marangoni stresses. Because $d\sigma/dc$ remains approximately constant over 1--100~mM concentration range \cite{pegram2007hofmeister}, the magnitude of $\Delta\sigma$ increases linearly with concentration, $c$, leading to progressively stronger Marangoni stresses.  Therefore, as the film drains, it is expected that in the limiting case, the inward Marangoni stress may render the interface immobile thus reducing the drainage rate with an increase in concentration.

The above difference in surface tension developed at the bubble-liquid interface in the thin film has been the basis for the conclusions in the literature that suggest that delay in coalescence is solely driven by the Marangoni stresses. Here, we show that although Marangoni stresses may lead to an immobile interface, but it is not sufficient to explain the 1--2 orders of magnitude delay in the coalescence time observed for electrolyte solutions. We propose that the primary reason for the delayed coalescence is the geometry of the rim, specifically the thickness of the film at the rim which is governed by EDL screening due to the presence of electrolytes. In order to delineate the effects of geometry from the boundary conditions imposed at the bubble-liquid interface, we estimate the drainage time for thin films of different geometries, namely, flat, linear, and double-welled, close to the one observed in experiments (see Supplementary Sec. S8) and boundary conditions. In these calculations, we keep the driving force (buoyancy or otherwise) the same for bubble-substrate interaction  for all geometries with the same $r_{rim}$. The temporal evolution equation for $h_{max}$ (Eq. S59) is solved numerically to obtain the variation in $h_{max}$ and the corresponding $h_{avg}$ with time; both $h_{max}$ and $h_{avg}$ are scaled by the initial average film thickness.
Figure S10 shows the evolution of the average film thickness indicating the drainage dynamics. For a flat film, the no-slip drainage is slower as compared to the corresponding evolution when slip boundary condition is imposed at the 
bubble-liquid interface. The ratio of the time taken to arrive at $h_{avg} = 0.2$ for no-slip to slip boundary condition for flat film is $\sim 3.5$.  However, for a double-welled interface, drainage is significantly slower even under slip boundary conditions compared to the flat film case with no-slip boundary condition.
 The ratio of times required to reach $h_{avg} = 0.9$ for double-welled and flat film under no-slip conditions is $\sim  3.2\times 10^3$, suggesting that the drainage time depends much more strongly on the film geometry than on the boundary conditions. 

 The above illustration, based on the simplified model of the role of rim geometry on the delay in drainage is further supported by experimental observations in the low bubble approach speed regime where dimple formation is absent \cite{hofmeier1995speed, yaminsky2010stability}. Under such conditions, the electrolyte concentration has been reported to have negligible effect on the drainage times. 

\begin{figure}
\centering
\includegraphics[width=1\linewidth]{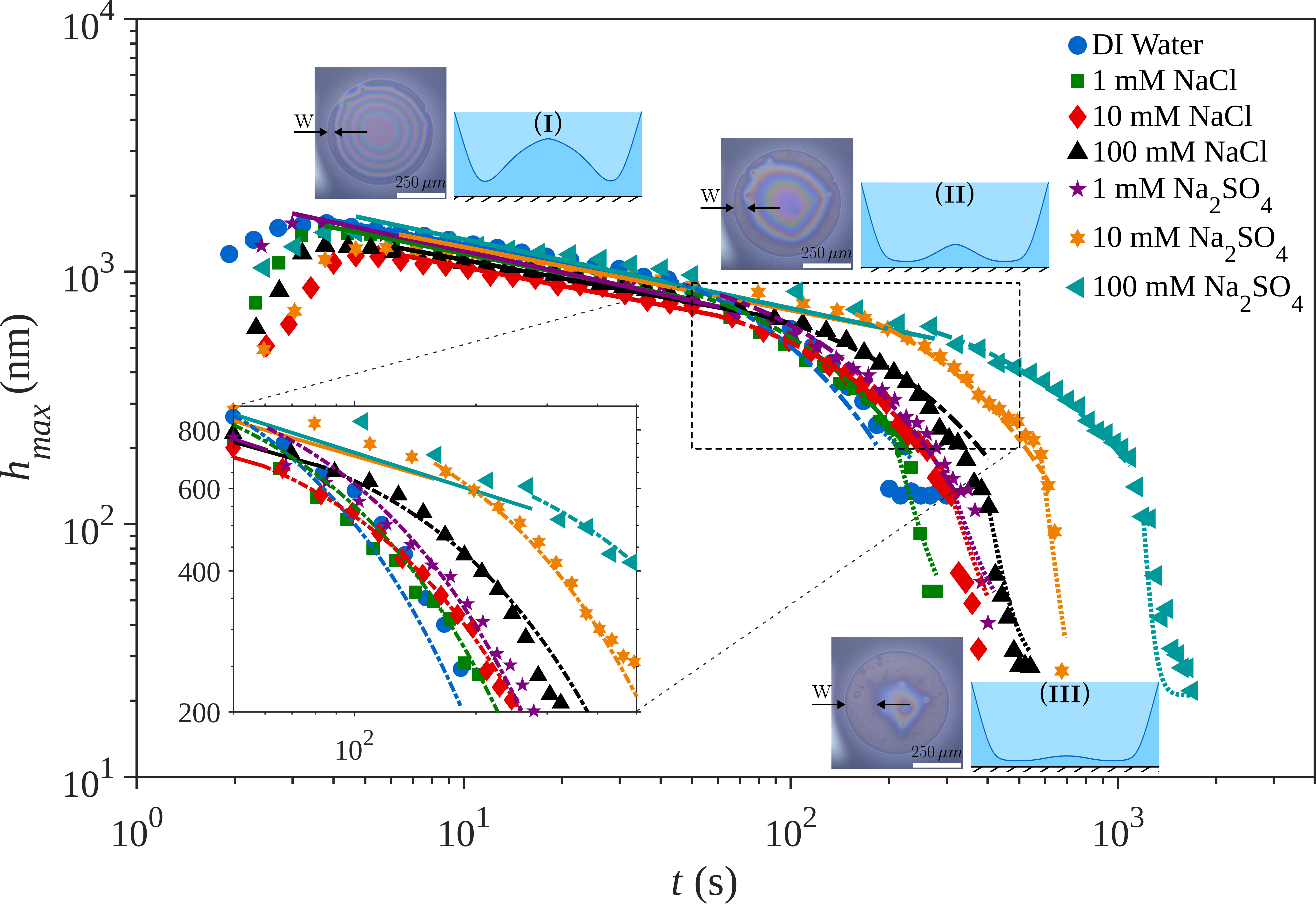}
\caption{Mechanistic interpretation and quantitative characterization of regime-dependent thin-film drainage. Temporal evolution of the maximum film thickness, $h_{\max}$, for DI water and electrolyte solutions of NaCl and Na$_2$SO$_4$ at various concentrations. The data exhibit three corresponding regimes: an early-time power-law scaling $h_{\max} \sim t^{-1/4}$ (Regime~I, solid lines), followed by a primary exponential relaxation $h_{\max} \sim e^{-t/\tau_1}$ (Regime~II, dash-dot lines), and a terminal exponential decay $h_{\max} \sim e^{-t/\tau_2}$ (Regime~III, dashed lines) as the film approaches equilibrium. The inset highlights the transition from the visco-capillary to the primary exponential regime on a log-log scale. Schematics in the inset show the representative film profiles for the three regimes along with the interferograms.}
\label{fig_Capillary_Vs_Marangoni}
\end{figure}

\subsection*{Scaling Analysis and Universal Drainage Dynamics}
\label{sec_Scaling}

While the experimental observations highlight the severe retardation of film drainage, it remains an open question whether the underlying kinematics follow a universal physical law. Figure~\ref{fig_Capillary_Vs_Marangoni} presents the temporal evolution of $h_{\max}$ for DI water, NaCl ($1, 10, 100~\mathrm{mM}$), and Na$_2$SO$_4$ ($1, 10, 100~\mathrm{mM}$). The data, irrespective of the salt type and concentration used, reveal three fundamental sequence, namely, Regime-I, Regime-II and Regime-III, in the drainage kinetics: an initial algebraic visco-capillary stage ($h_{\max} \sim t^{-1/4}$), a subsequent primary exponential relaxation ($h_{\max} \sim \exp(-t/\tau_1)$), and a terminal, secondary exponential decay ($h_{\max} \sim \exp(-t/\tau_2)$), where $\tau_1$ and $\tau_2$ are the salt and concentration dependent characteristic relaxation times as discussed below. These successive crossovers mark critical shifts in the nature of drainage, separated by characteristic transition points as summarized in Table~\ref{tab_combined_scaling_parameters}.

\begin{table}[htpb]
\centering
\caption{Characteristic transition parameters and drainage timescales for DI water, NaCl, and Na$_2$SO$_4$ solutions at different concentrations. Here, $t_{0}$ and $h_{0}$ denote the transition time and thickness corresponding to the onset of Regime~II following Regime~I; $n_I$ is the drainage exponent in Regime~I ($t^{n_I}$); $\tau_1$ and $\tau_2$ are the primary and terminal relaxation times in Regimes~II and III, respectively.}
\renewcommand{\arraystretch}{1.25}
\begin{tabular}{l c @{\hspace{0.8cm}} c c c @{\hspace{0.8cm}} c c c}
\hline\hline
\multirow{2}{*}{Parameter}
& \multirow{2}{*}{DI Water}
& \multicolumn{3}{c}{NaCl}
& \multicolumn{3}{c}{Na$_2$SO$_4$} \\
\cline{3-5} \cline{6-8}
&
& 1 mM & 10 mM & 100 mM
& 1 mM & 10 mM & 100 mM \\
\hline
$t_{0}$ (s)
& 40
& 47 & 58 & 80
& 60 & 157 & 275 \\

$h_{0}$ (nm)
& 959
& 834 & 673 & 675
& 808 & 680 & 577 \\

$n_I$
& -0.237
& -0.258 & -0.247 & -0.241
& -0.263 & -0.250 & -0.273 \\

$\tau_1$ (s)
& 92.84
& 125.00 & 164.01 & 245.01
& 142.70 & 298.56 & 663.01 \\

$\tau_2$ (s)
& 68.47
& 41.21 & 64.91 & 44.87
& 59.60 & 71.55 & 53.64 \\
\hline\hline
\end{tabular}
\label{tab_combined_scaling_parameters}
\end{table}

\subsubsection*{Regime I: Visco-Capillary Dimple Formation ($h_{\max} \sim t^{-1/4}$)}

In the initial stage of drainage, the film evolves under the combined action of capillary forces and viscous resistance, giving rise to a dimple profile, defined by a central bulge and rim height, that govern the drainage dynamics characterizing Regime~I. In this regime, the rim width remains small, indicating that the hydraulic resistance is highly localized within a narrow rim region. 

Approximating the film shape by using a downward facing parabolic profile for dimple region and an upward cubic profile (retaining third order terms in the Taylor series expansion) for the rim region, we obtain the relation between $h_{min}$ and $h_{max}$ by performing an asymptotic matching of the two profiles and mass conservation (see Sec. S5 in the Supplemental Material):
\begin{equation}
h_{\min} = \frac{27 \mu r_{rim}^3}{4 \sigma h_{max}^3} q(r_{rim}),
\label{eq:hmin_qb_relation}
\end{equation}
where $q(r_{rim})$ is the integrated volumetric flux at the rim. Integrated momentum balance yields:
\begin{equation}
    \frac{2 \sigma}{R_L}   = \frac{12 q(r_{rim}) \mu}{(3 \beta - 2) h_{min}^3}(r_{rim} - r_c)
    \label{eq:dhdtRegimeI}
\end{equation}
Here, $r_c$ is the radial location for matching the dimple and rim profiles given by:
\begin{equation}
r_{rim} - r_c =  \frac{3 h_{min} r_{rim}}{2 h_{max}(t)}.
\label{eq:exp-rc}
\end{equation}.
Using the above relation with $h_{min}$ from Eq.\ref{eq:hmin_qb_relation} and $q(r_{rim}) \sim -h_{max}'(t) r_{rim}/4$ (see Supplementary Sec S5.1) gives:
\begin{equation}
  \frac{1}{h_{max}^5}\frac{dh_{max}}{dt} =  -\frac{64(3 \beta - 2)}{81}\frac{\sigma R_L}{r_{rim}^6 \mu}
\end{equation}
which results in the scaling:
\begin{equation}
h_{max}(t) \simeq \frac{3}{4} \left( \frac{\mu r_{rim}^6}{(3\beta -2)\sigma R_L} \right)^{1/4} t^{-1/4}.
\end{equation} 

In the above scaling for $h_{max}$, the parameter $\beta$ characterizes the interfacial velocity boundary condition at the air--water interface, with $\beta = 1$ corresponding to an immobile interface and $\beta = 2$ representing a shear-free interface. The characteristic exponent $-1/4$ arises from the coupled visco-capillary drainage of a dimpled thin film with a confined rim geometry. Accordingly, all investigated systems, including DI water, exhibit a power-law drainage behavior with an exponent $n_{\mathrm{I}} \approx -1/4$ (Table~\ref{tab_combined_scaling_parameters}), consistent with classical visco-capillary thinning in dimpled thin films \citep{hartland1977model, frankel1962dimpling, bluteau2017water}.
Regime-I continues till film thickness at the rim reaches $h_{stable}$. Since, for higher concentration of salts, $h_{stable}$ is smaller (see Table \ref{tab_combined_scaling_parameters}), Regime-I is extended further.

The experimentally observed early-stage dynamics for all the solutions are found to be consistent with a nearly shear-free interface ($\beta \approx 2$), indicating that the air--water interface remains largely mobile throughout Regime~I. 

\subsubsection*{Regime II: Visco-EDL Dimple Flattening ($h_{\max} \sim \exp(-t/\tau_1)$)}

As drainage progresses ($h_{max} < h_0$ and $t > t_0$; see Table \ref{tab_combined_scaling_parameters}), the nature of the temporal evolution of the film thickness changes (see Figure \ref{fig_Capillary_Vs_Marangoni}). Once the rim thins ($\sim h_{stable}$; see Table~\ref{tab_stabilised_para_combined}), the repulsive EDL pressure  balances the driving capillary pressure at the rim. As detailed in Section~S5 of the Supplementary Information, this force balance results in a locally flat rim with a fixed equilibrium thickness ($h_{\min} = h_{\mathrm{stable}}$), marking the onset of Regime~II. Concurrently, the rim undergoes geometric flattening, leading to an increase in the effective rim width while the outer rim radius remains unchanged (see inset for Regime-II in Figure~\ref{fig_Capillary_Vs_Marangoni}). This widening of the rim region results in a transition from a highly localized flow resistance at the rim to a more spatially extended hydraulic bottleneck.

Assuming $h_{min} = h_{stable}$ at the rim, the evolution equation (Eq.\ref{eq:dhdtRegimeI}) can be written as:
\begin{equation}
   \frac{2\sigma}{R_L} \sim -\frac{18 r_{rim}^2 h_{max}'(t)\mu}{4(3 \beta - 2)h_{stable}^2 h_{max}(t)}.
\end{equation}
resulting in a departure of the central film thickness from the power law scaling to an exponential evolution:
\begin{equation}
    h_{\max}(t) \propto \exp(-t/\tau_1),
    \label{eq:hmaxvst-RII}
\end{equation}
where the characteristic time constant, $\tau_1$, is given by,
\begin{equation}
    \tau_1 = \frac{9 R_L \mu r_{\mathrm{rim}}^2}{16 \sigma h_{\mathrm{stable}}^2}.
\end{equation}
Thus, the exponential decrease of the film thickness in Regime II is governed by a time constant dictated by the fluid viscosity ($\mu$), surface tension ($\sigma$), and the disjoining-pressure-stabilized film thickness ($h_{\mathrm{stable}}$). 

The experimentally extracted time constants, $\tau_1$, for all the investigated systems are summarized in Table~\ref{tab_combined_scaling_parameters}. For DI water, $\tau_1$ is relatively small, $\tau_1 \approx 92.8~\mathrm{s}$ and increases monotonically with electrolyte concentration $\tau_1 \approx 245.0~\mathrm{s}$ for $100~\mathrm{mM}$ NaCl. The retardation of drainage is further amplified by the ion valency reaching $\tau_1 \approx 663.0~\mathrm{s}$ for $100~\mathrm{mM}$ Na$_2$SO$_4$. We note that $\tau_1$ is dictated by the inverse-square dependence on the equilibrium thickness ($\tau_1 \propto h_{\mathrm{stable}}^{-2}$). Thus, even a slight decrease in $h_{stable}$ significantly increases $\tau_1$. As ionic strength and valency increase, enhanced EDL screening compresses the Debye length, reducing the rim film thickness to a significantly thinner $h_{\mathrm{stable}}$ (see Table~\ref{tab_combined_scaling_parameters}) which creates a severe hydraulic bottleneck, increasing the viscous resistance to radial flow and thereby prolonging the drainage of the film. The $\tau_1$ obtained theoretically, results in an excellent agreement between the experimental observations and the theoretically predicted film evolution.

\subsubsection*{Regime III: EDL-driven uniform relaxation to equilibrium film thickness ($h_{\max} \sim \exp(-t/\tau_2)$)}

As the dimple in the central region of the film completely relaxes, the drainage enters Regime~III, characterized by a nearly uniform, thin planar geometry ($h \approx h_{\mathrm{stable}}$). At this stage, the rim region widens almost up to the central region (see 
regime III inset in Fig.~\ref{fig_Capillary_Vs_Marangoni}). In contrast to the localized rim in Regime~I and the moderately spread rim in Regime~II, the flow in this regime is  reflective of the drainage arising from fully developed radial flow between nearly parallel spatial confinement. At these extreme nanometric confinements, the repulsive electrical double layer (EDL) disjoining pressure extends across the entire film domain resisting drainage.

The film evolution in this regime is determined by the balance between the capillary pressure ($2\sigma/R_L$) that drives the drainage and the EDL disjoining pressure ($\Pi_{EDL} = K e^{-h/\lambda}$), and the viscous hydrodynamic resistance. Normal stress balance at the bubble-liquid interface, and the overall pressure drop across the film due to viscous forces, results in:
\begin{equation}
    \frac{2\sigma}{R_L} - K e^{-h/\lambda} = -\frac{12\mu R_c^2}{4(3\beta-2)h^3} \frac{dh}{dt},
    \label{eq:regime3-geq}
\end{equation}
where $K = 64 k_B T n_{\infty} \tanh^2[(z_i e\psi_0)/(4 k_B T)]$ is the electrostatic interaction parameter and $R_c$ is the fixed outer radius of the rim.

As the draining film flattens toward its equilibrium thickness ($h \to h_{\mathrm{stable}}$), the drainage kinematics asymptotically seizes, satisfying the eventual force balance $2\sigma/R_L = K e^{-h_{\mathrm{stable}}/\lambda}$, giving $h_{stable} = \lambda ln(K R_L/2\sigma)$. To analyze the late-stage drainage dynamics approaching this limit, we introduce a small perturbation thickness, $\tilde{h}(t)$, representing the  deviation from the equilibrium:
\begin{equation}
    h(t) = h_{\mathrm{stable}} + \tilde{h}(t).
\end{equation}
Assuming, $\tilde{h} \ll h_{\mathrm{stable}}$ and $\tilde{h} \ll \lambda$, in this regime, we employ the approximations: $e^{-\tilde{h}/\lambda} \approx 1 - \tilde{h}/\lambda$ and  $(1 + \tilde{h}/h_{\mathrm{stable}})^3 \approx 1$. Substituting the perturbation into Eq.\ref{eq:regime3-geq}, yields a first-order linear ordinary differential equation (as detailed in Section S5 of the Supplementary Information):
\begin{equation}
    \frac{d\tilde{h}}{dt} = - \left( \frac{8(3\beta-2)\sigma h_{\mathrm{stable}}^3}{12\mu R_c^2 R_L \lambda} \right) \tilde{h}.
\end{equation}
Integrating, we get an exponential relaxation towards the stable film thickness:
\begin{equation}
    h(t) = h_{\mathrm{stable}} + \tilde{h}_0 \exp\left( - \frac{t}{\tau_2} \right),
\end{equation}
where $\tilde{h}_0$ is the initial perturbation thickness at the onset of Regime~III, and the terminal relaxation time, $\tau_2$, is defined as:
\begin{equation}
    \tau_2 = \frac{3\mu R_c^2 R_L \lambda}{8\sigma h_{\mathrm{stable}}^3}.
\end{equation}

Interestingly, the time constant, $\tau_1$ and $\tau_2$ show the same dependence on $\lambda^{-2}$. Values for $\tau_2$ computed for all solutions are summarized in Table~\ref{tab_combined_scaling_parameters}. Theoretical values of $\tau_2$ predict the experimental film drainage accurately, establishing the underlying mechanism of drainage in this regime.

\subsubsection*{Universal Scaling}

\begin{figure}
    \centering
    \includegraphics[width=0.9\linewidth]{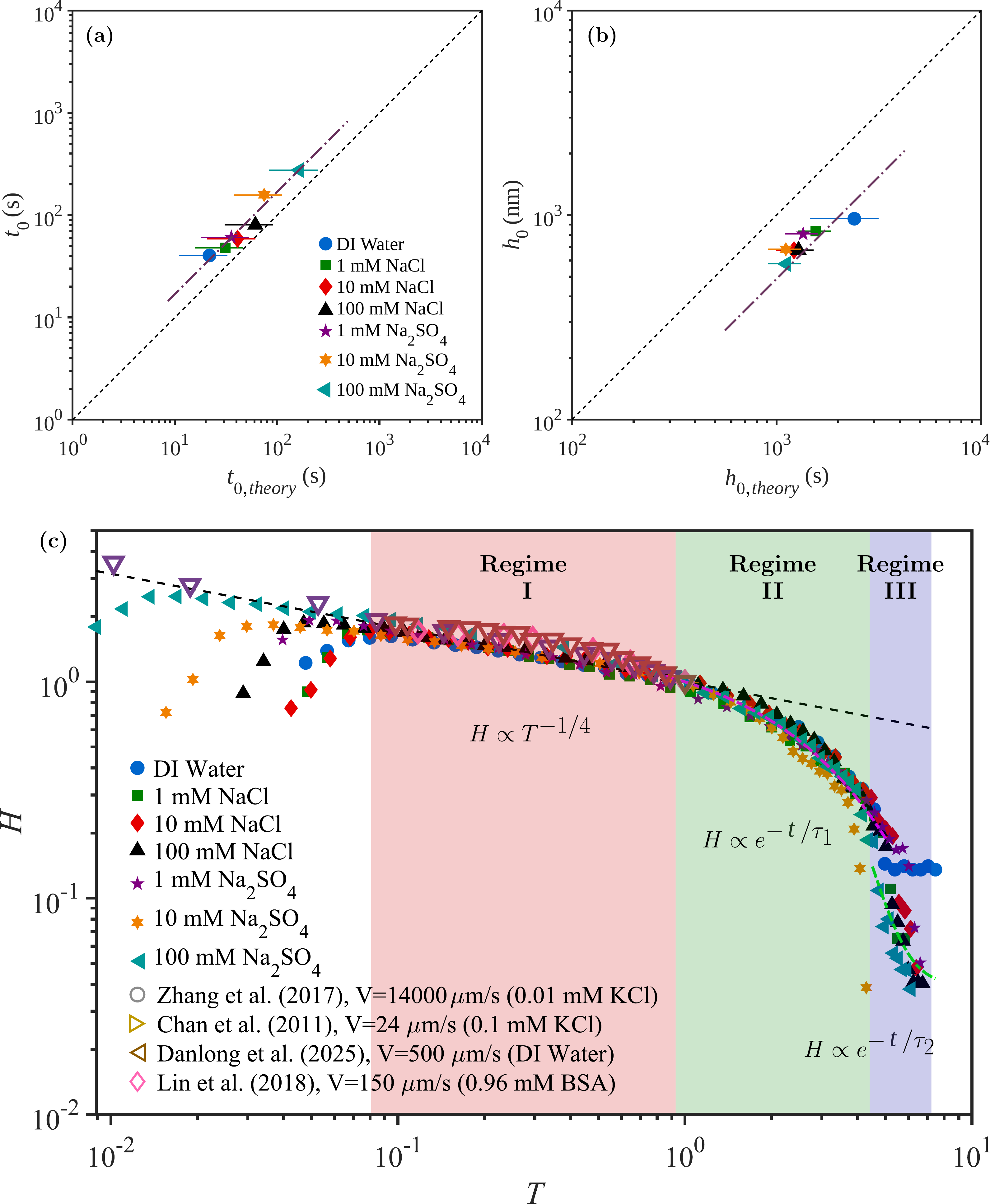} 
    \caption{Comparison between experimentally measured and theoretically predicted transition scales, together with the universal scaling of thin-film drainage dynamics. Panels (a) and (b) compare the experimentally determined transition time ($t_{\mathrm{exp}}$) and transition thickness ($h_{\mathrm{exp}}$) with the corresponding theoretical predictions ($t_{\mathrm{theory}}$, $h_{\mathrm{theory}}$) for DI water, NaCl, and Na$_2$SO$_4$ solutions at different concentrations. The dashed line represents perfect agreement. 
    Panel (c) shows the universal scaling collapse of the dimensionless film thickness, $H = h_{\max}/h_0$, as a function of the dimensionless time, $T = t/t_0$, where $(t_0,h_0)$ are the experimentally determined transition scales between Regimes~I and II. Despite variations in electrolyte type, concentration, and ion valency, all datasets collapse onto a single master curve exhibiting three successive drainage regimes: an initial visco-capillary regime ($H \propto T^{-1/4}$), an intermediate exponential relaxation ($H \propto e^{-t/\tau_1}$), and a terminal nanoscale exponential decay ($H \propto e^{-t/\tau_2}$). Literature data also follow the same early-time scaling, confirming the universality of the visco-capillary drainage dynamics across diverse systems and experimental conditions.}
    \label{fig_NaClNa2SO4hmax_slope}
\end{figure}

Figures~\ref{fig_NaClNa2SO4hmax_slope}(a,b) compare the experimentally measured transition scales $(t_0,h_{0})$ with the corresponding theoretical predictions $(t_{0,\mathrm{theory}},h_{0,\mathrm{theory}})$, (derivation shown in Section S6 of Supplementary Information),  where, 

\begin{equation}
t_{0,\mathrm{theory}} = \frac{\tau_1}{4},
\qquad
h_{0,\mathrm{theory}} = \sqrt{\frac{3}{2}}\,r_{\mathrm{rim}}\sqrt{\frac{h_{\mathrm{stable}}}{R_L}}.
\end{equation}
The model predicts the transition times reasonably well across all investigated systems, indicating that the onset of Regime II is governed primarily by rim-controlled hydrodynamic resistance. In contrast, the transition thickness is slightly overpredicted. This discrepancy arises because the transition criterion is satisfied locally when the rim thickness reaches its equilibrium value, $h_{\min}=h_{\mathrm{stable}}$, whereas the central dimple thickness, $h_{\max}$, continues to evolve over a finite timescale before fully responding to the onset of disjoining-pressure effects. Consequently, the transition time is captured accurately by the rim-based criterion, while the corresponding transition thickness reflects a delayed geometric response of the dimple.

Despite pronounced variations in the absolute timescales of entrapped-film drainage arising from changes in ionic strength and surface activity, the underlying drainage dynamics and the transitions between the three successive regimes remain self-similar. The evolution of the dimple thickness, $h_{\max}$ can be collapsed onto a single master curve when scaled by characteristic transition values, $t_0, h_0$, (see Table~\ref{tab_combined_scaling_parameters}) as shown in Figure~\ref{fig_NaClNa2SO4hmax_slope}(c). These transition times span nearly an order of magnitude, from $t_{0}\approx 40~\mathrm{s}$ at low electrolyte concentrations to $t_{0}\approx 275~\mathrm{s}$ for the $100~\mathrm{mM}$ Na$_2$SO$_4$ solution, highlighting the strong influence of interfacial retardation on the drainage dynamics. 

The universality indicates that, despite substantial differences in electrolyte type, concentration, and ion valency, the three fundamental drainage sequence, an initial phase of visco-capillary thinning, followed by a primary exponential geometric relaxation, and ultimately a terminal nanoscale exponential decay, remains unchanged. Notably, the early-time dynamics in Regime~I show excellent agreement with previously reported experimental data from the literature, as also highlighted in Figure~\ref{fig_NaClNa2SO4hmax_slope}(c), confirming the robustness of the visco-capillary scaling. Additional comparisons with a broader range of literature data are provided in Supplementary Figure~S9. Rather than modifying the nature of the drainage regimes themselves, the presence of electrolytes primarily delays the transitions between them. Thus, electrolyte-specific properties therefore determine the characteristic transition scales $(t_0,h_0)$, while the underlying hydrodynamic pathway governing the drainage remains universal.

To summarize, this study provides conclusive evidence for the existence of three distinct drainage regimes during thin-film drainage in bubble--substrate interactions. These regimes are observed consistently for DI water as well as NaCl and Na$_2$SO$_4$ electrolyte solutions over a broad range of concentrations, demonstrating that the three-stage drainage sequence is a generic feature of the process rather than a system-specific phenomenon.

By combining high-resolution interferometric measurements with theoretical analysis, we identify the physical mechanisms governing each regime: visco-capillary drainage of the entrapped liquid film, rim-flattening-induced exponential drainage, and electrostatic double-layer (EDL) mediated exponential relaxation towards a stable thin-film state. An asymptotic matching analysis yields the characteristic transition scales separating the different drainage stages and provides a quantitative framework for predicting the onset of the delayed-drainage regime.

Most importantly, we show that when the drainage dynamics are scaled using these transition scales, all datasets collapse onto a single master curve. This collapse is observed not only for the experiments performed in the present study but also for drainage data reported previously in the literature. Despite substantial variations in electrolyte concentration, ionic strength, ion valency, and surface activity, the drainage dynamics and the transitions between the three successive regimes remain self-similar. These results reveal a universal hydrodynamic pathway governing thin-film drainage during bubble--substrate interactions and establish a unifying framework for understanding delayed coalescence phenomena across a broad class of fluid interfaces.

\section*{Methods}\label{sec_Experimental_Setup}

\subsection*{Experimental setup and color interferometry}

Figure~\ref{fig_ExpSetup_Fringes}(a) shows the experimental arrangement used to study the approach and interaction of an air bubble with a flat glass substrate in an electrolyte medium. The drainage dynamics of the intervening liquid film is analyzed via color interferometry. As the bubble approaches the substrate, the intervening liquid film gradually thins. Once the film thickness becomes comparable to the coherence length of the illuminating light source, interference fringes emerge due to the optical path difference between the light reflected from the air--liquid and liquid--solid interfaces..  To capture these dynamic interference patterns, an upright microscope (Leica M205 A) fitted with a 5$\times$ objective and a high-definition color camera (Leica MC 190 HD) is used. The images are recorded at 100 frames per second and a spatial resolution of $0.83~\mu\mathrm{m}/\mathrm{pixel}$. Illumination is provided by a white-light LED source (Leica KL2500) with spectral emissivity well-matched to the camera sensitivity. 
Simultaneously, the macroscopic position and size of the bubble are recorded from a side view using a high-speed camera (Photron FASTCAM Mini AX200) operating at 2000 frames per second. The camera is equipped with a 6$\times$ long-working-distance lens (Navitar) and back-lit by an LED panel (Nila Zaila), yielding a spatial resolution of $11.23~\mu\mathrm{m}/\mathrm{pixel}$

Air bubbles of radius, $R = 1.39 \pm 0.1$~mm are generated at the tip of a stainless-steel capillary (i.d.\ 1.54~mm and o.d.\ 1.83~mm; Nordson EFD) immersed in a cuvette (75~mm $\times$ 50~mm $\times$ 50~mm) filled with electrolyte solutions. The solutions are prepared by dissolving analytical-grade NaCl and Na$_2$SO$_4$ salts (Sigma-Aldrich) in deionized water (initial conductivity 128.8~$\mu$S~cm$^{-1}$). Bubble growth on the capillary is controlled using a syringe pump (Chemix Fusion 200) and a precision flow valve (Darwin Microfluidics, LVF-KMM-08) connected in series with the syringe pump and capillary. A glass slide (76~mm $\times$ 26~mm $\times$ 1~mm; BOROSIL Scientific, 9100P02) is placed in a recessed groove at the top of the cuvette and lightly loaded to prevent movement of the glass slide. The initial separation between the bubble apex and the glass--electrolyte interface is $h_{\mathrm{ini}} = 601 \pm 5~\mu$m. The bubble is then translated towards the glass at a constant velocity $V_b = 350 \pm 2~\mu\text{m}~\text{s}^{-1}$ over a displacement $\Delta h = 751 \pm 5~\mu\text{m}$ using a microcontroller-driven $z$-stage (Holmarc MVTS-75-25), after which it is allowed to evolve freely. All experiments are conducted under standard ambient temperature and pressure conditions.

\subsection*{Film thickness reconstruction}

Quantitative film-thickness profiles are reconstructed from the recorded color interferograms using a highly calibrated color-matching procedure \cite{van2012direct}. Briefly, experimental sRGB interference patterns are converted into the CIE 1976 CIELAB color space, which effectively decouples the illumination intensity from the true chromatic components. The local film thickness is then determined by matching these chromatic coordinates against a high-resolution calibration database, which is independently generated using a reference plano–convex lens of known geometry. 

Because interference colors are periodic, this matching process yields a localized color-difference field containing multiple candidate profiles. The final, physically admissible interface profile is extracted by enforcing spatial smoothness and minimizing the global color deviation. This rigorous methodology yields an absolute thickness accuracy of $\pm 18$~nm and a typical measurement uncertainty of $\pm 14$~nm. A comprehensive description of the optical calibration, signal processing, and profile-selection algorithms is detailed in Section S1 of the Supplementary Information.

\section*{Data availability}
The datasets generated and/or analyzed during the current study are available from the corresponding author on reasonable request.

\section*{Code availability}
The custom MATLAB scripts utilized for image processing, color interferometry calibration, and thickness reconstruction are available from the corresponding author on reasonable request.


\vspace{2cm}
\begin{center}
    \textbf{\Large Supplementary Information}
\end{center}

\setcounter{section}{0}
\setcounter{figure}{0}
\setcounter{table}{0}
\setcounter{equation}{0}

\renewcommand{\thesection}{S\arabic{section}}
\renewcommand{\thefigure}{S\arabic{figure}}
\renewcommand{\thetable}{S\arabic{table}}
\renewcommand{\theequation}{S\arabic{equation}}


\section{Color Interferometry and Film Thickness Reconstruction}\label{sec_ColorInterferometry}

\begin{figure}[h!]
    \centering
    \includegraphics[width=0.75\linewidth]{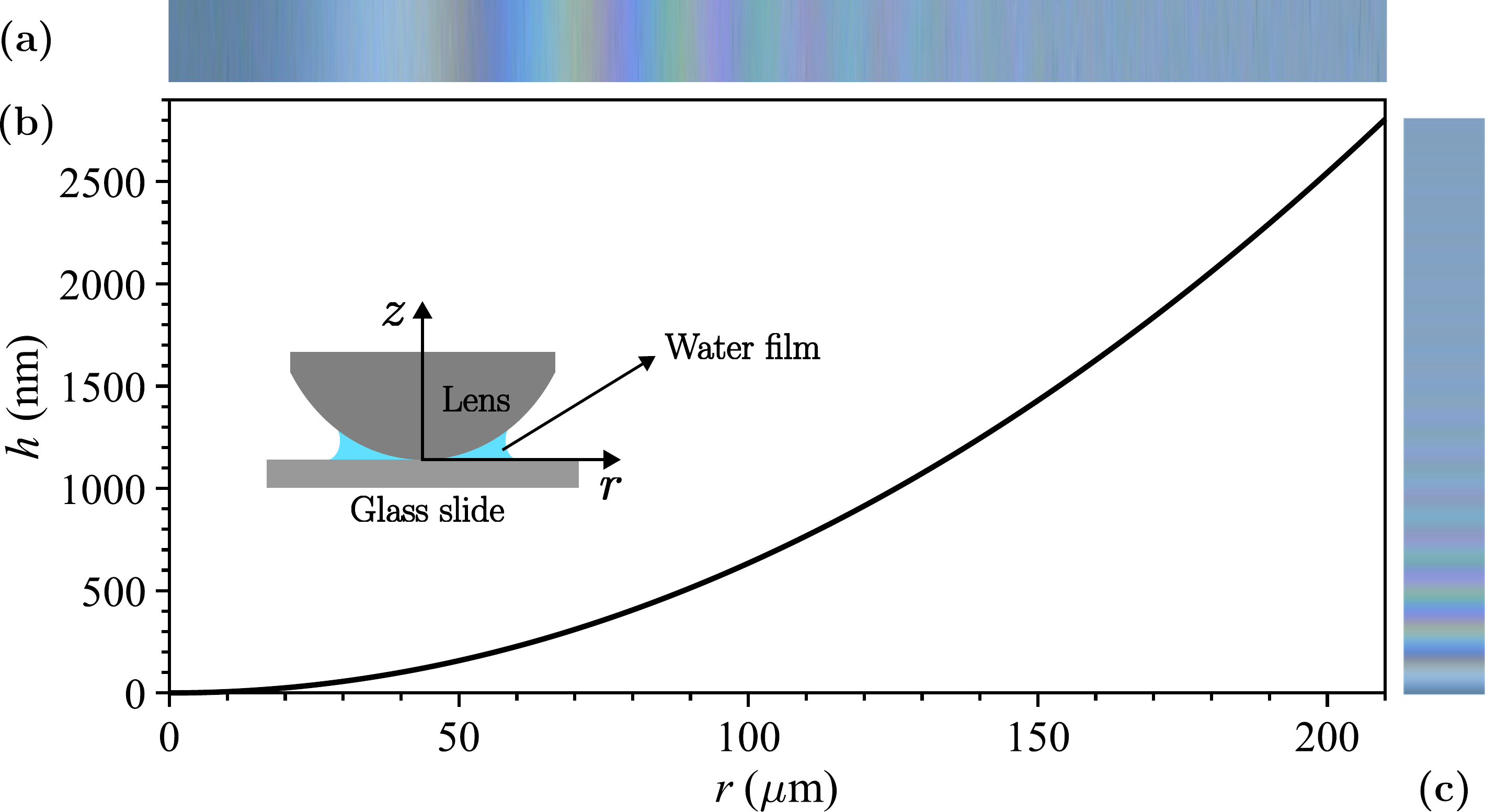}
    \caption{Schematic of the reference interferometric arrangement: a plano–convex lens of known radius of curvature ($7.85$ mm) placed over a glass slide with a thin DI-water film in between. The resulting interference pattern provides a calibrated correspondence between color and film thickness.}
    \label{fig_RefLensArrngmnt}
\end{figure}

The liquid film thickness between the bubble and the substrate is measured using color interferometry following \cite{van2012direct}. A calibration database is first constructed using a reference interferometric configuration comprising a plano–convex lens (Edmund Optics, $R = 7.85$ mm), a thin film of deionized water (conductivity $128.8~\mu$S~cm$^{-1}$), and a planar glass substrate (Borosil Microslides), as illustrated in Figure~\ref{fig_RefLensArrngmnt}.

The observed reference interference pattern originates from the phase difference between light reflected at the lens–water and water–glass interfaces. Owing to the known geometry of the lens, the film thickness varies radially as
\begin{equation}
    h(r) = R - \sqrt{R^2 - r^2},
\end{equation}
thereby providing a direct mapping between interference color and absolute film thickness.

To process the experimental data, a horizontal strip (50 pixels wide) is first extracted from the interferogram and averaged across its width. This spatial averaging suppresses random sensor noise and significantly enhances the signal-to-noise ratio of the intensity profile. To ensure robustness against illumination variations, the averaged sRGB arrays are converted to the CIE 1976 (CIELAB) color space \cite{van2012direct}, where luminance ($L^*$) is decoupled from the chromatic components ($a^*$, $b^*$). The calibration database is thus constructed in terms of $(a^*, b^*)$ as a function of thickness.

\begin{figure}[h!]
    \centering
    \includegraphics[width=0.95\linewidth]{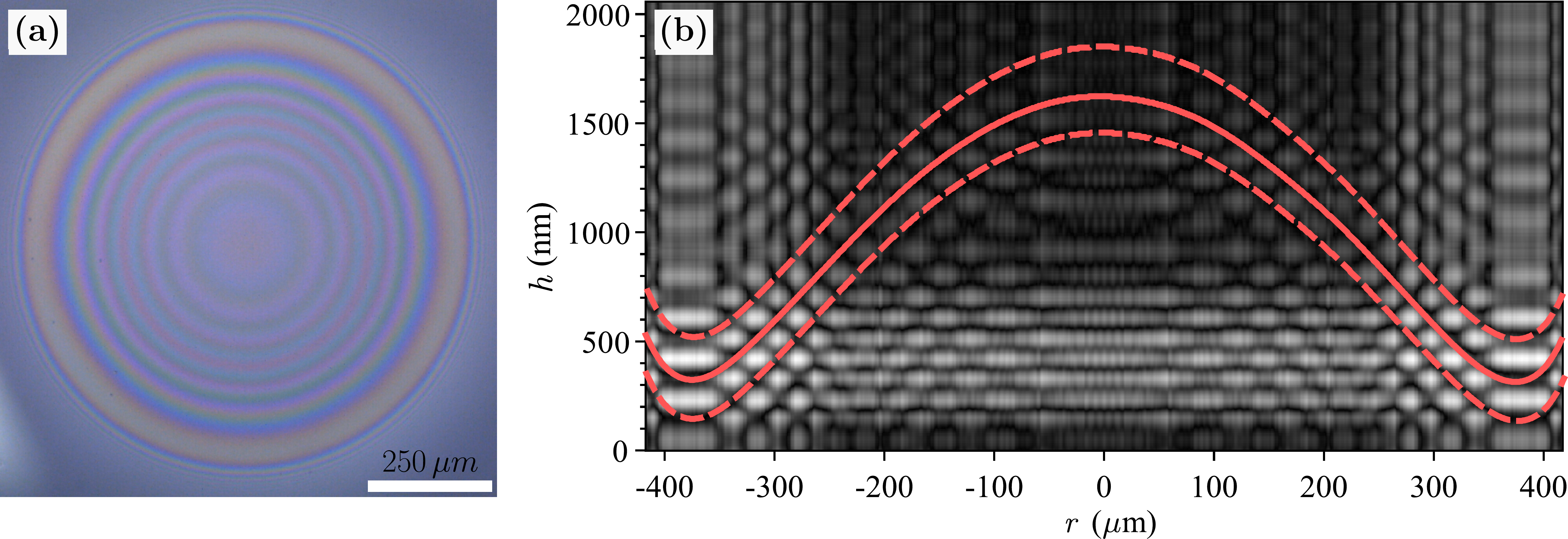}
    \caption{Extraction of thin-film profiles from color interferometry during bubble--glass interaction. (a) Representative color interferogram recorded during film drainage in deionized water, showing concentric interference fringes corresponding to variations in film thickness. (b) Grayscale color-difference image obtained by subtracting the interferogram from a reference pattern, highlighting fringe contrast for quantitative analysis. Superimposed curves represent candidate interface profiles, with the solid line denoting the selected profile used for thickness reconstruction.}
    \label{fig_FringeNdEImage}
\end{figure}

For an experimental bubble interferogram (Figure~\ref{fig_FringeNdEImage}(a)), the film thickness is determined by computing the $L_2$ norm (Euclidean color distance) between the measured chromatic coordinates $(a_j^*, b_j^*)$ at each radial location $r_j$ and the calibrated database $(a_i^*, b_i^*)$:
\begin{equation}
    dE_{ij} = \sqrt{(a_j^* - a_i^*)^2 + (b_j^* - b_i^*)^2},
\end{equation}
where $i$ spans all discrete thickness values $h_i$. This generates a two-dimensional color-difference map in the $(r,h)$ parameter space, where local dark minima correspond to candidate thickness solutions.

\begin{figure}[h!]
    \centering
    \includegraphics[width=0.7\linewidth]{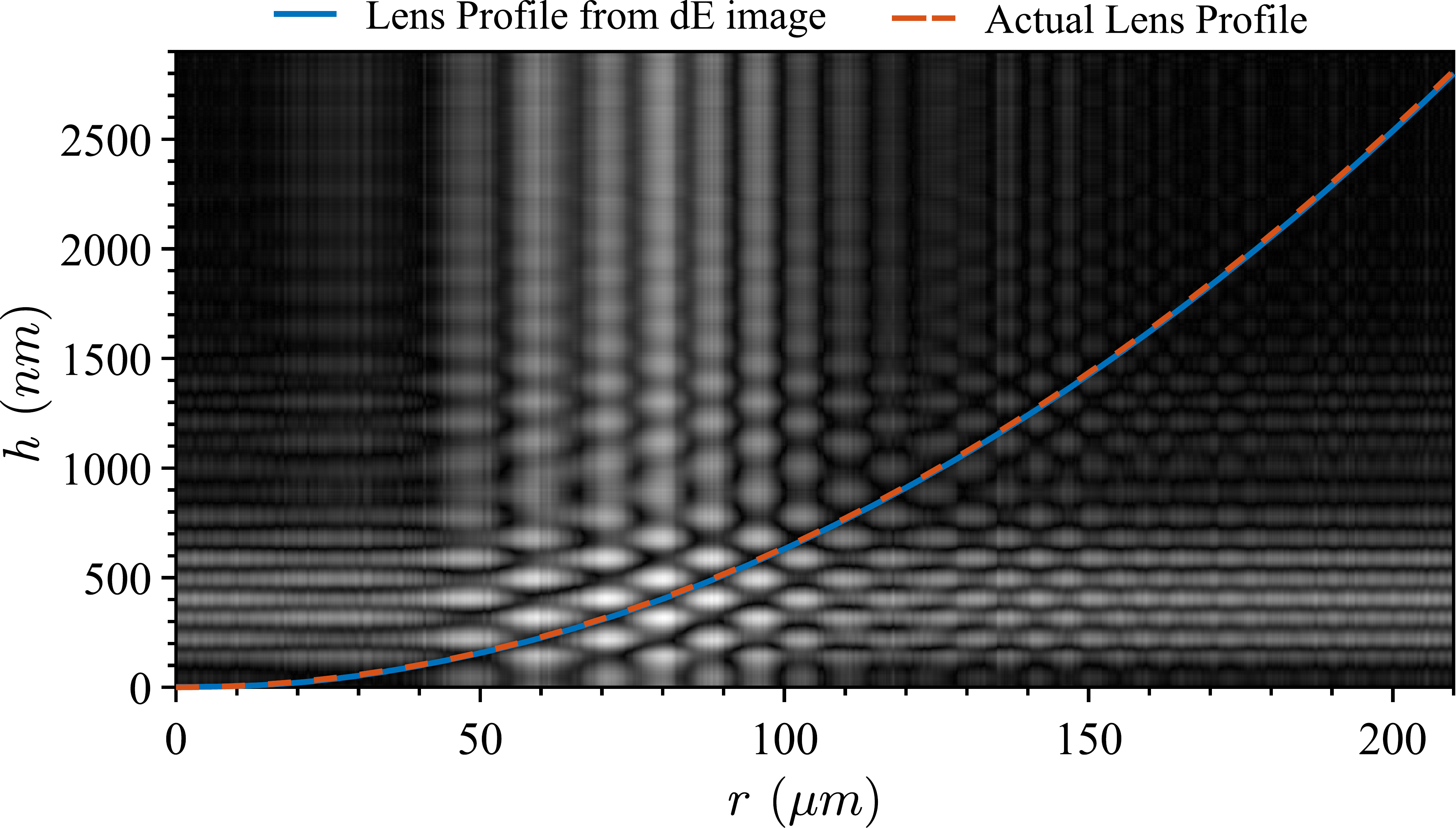}
    \caption{Validation of the color–interferometry calibration. The differential color-distance ($dE$) map from the reference lens configuration is overlaid with the reconstructed thickness profile (blue curve) and compared against the theoretical lens geometry (orange dashed line).}
    \label{fig_ExpMethdValidtn1}
\end{figure}

Due to the periodic nature of optical interference, a given color may correspond to multiple discrete thicknesses, leading to several visually similar candidate profiles (Figure~\ref{fig_FringeNdEImage}). The physically admissible solution is identified by enforcing spatial continuity and smoothness, and by selecting the continuous profile that minimizes the averaged color deviation along the profile:
\begin{equation}
    \langle dE \rangle_L = \frac{1}{N_L} \sum_{L} dE_{ij},
\end{equation}
where $N_L$ denotes the number of pixels along a candidate profile $L$. The continuous profile that yields the minimum $\langle dE \rangle_L$ is selected as the final solution (solid line in Figure~\ref{fig_FringeNdEImage}b), ensuring an accurate and physically consistent reconstruction of the spatio-temporal film profile $h(r,t)$.

The thickness resolution is governed by the sensitivity of chromatic variation with respect to film thickness within the calibrated range. The accuracy of the method is independently validated using the reference lens; as shown in Figure~\ref{fig_ExpMethdValidtn1}, the reconstructed profile exhibits excellent agreement with the theoretical geometry, providing an absolute thickness accuracy within $\pm 18$~nm. 

The measurement uncertainty is estimated from the finite width of the dark minima in the color-difference ($dE$) maps. This bandwidth arises due to the local sensitivity of the color-to-thickness gradient; in regions where the interference colors evolve gradually with thickness, the minima appear broader. Additionally, experimental factors such as light intensity fluctuations and the pixel-averaging techniques used to reduce noise contribute to this spatial dispersion. Thickness values extracted along the centers of these minima exhibit a small spread, and repeated measurements of these profiles yield a typical variation (uncertainty) of approximately $\pm 14$~nm.

\section{Interferograms of Thin-Film Evolution in NaCl and Na$_2$SO$_4$ Solutions}\label{Appdx_Na2SO4Evln.}

	\begin{figure}
		\centering		
        \includegraphics[width=\linewidth]{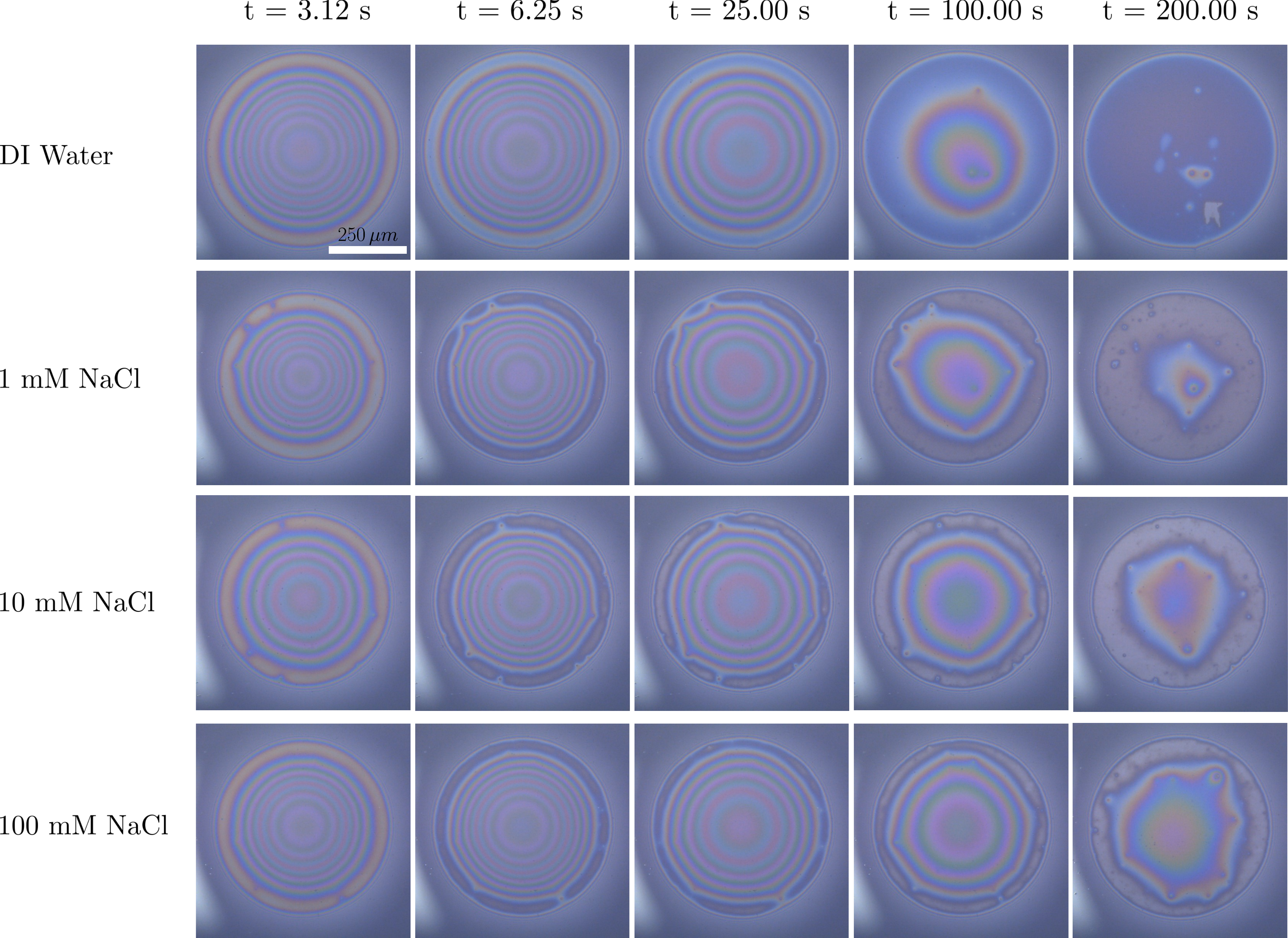}
        \caption{Color interferograms recorded at selected time instants during film drainage for deionised water and NaCl solutions of increasing concentration. Rows correspond to different electrolyte concentrations and columns to successive times after the onset of bubble approach.}
	\label{fig_DIWater_NaCl_AllConc}
	\end{figure}

\begin{figure}
    \centering		
    \includegraphics[width=\linewidth]{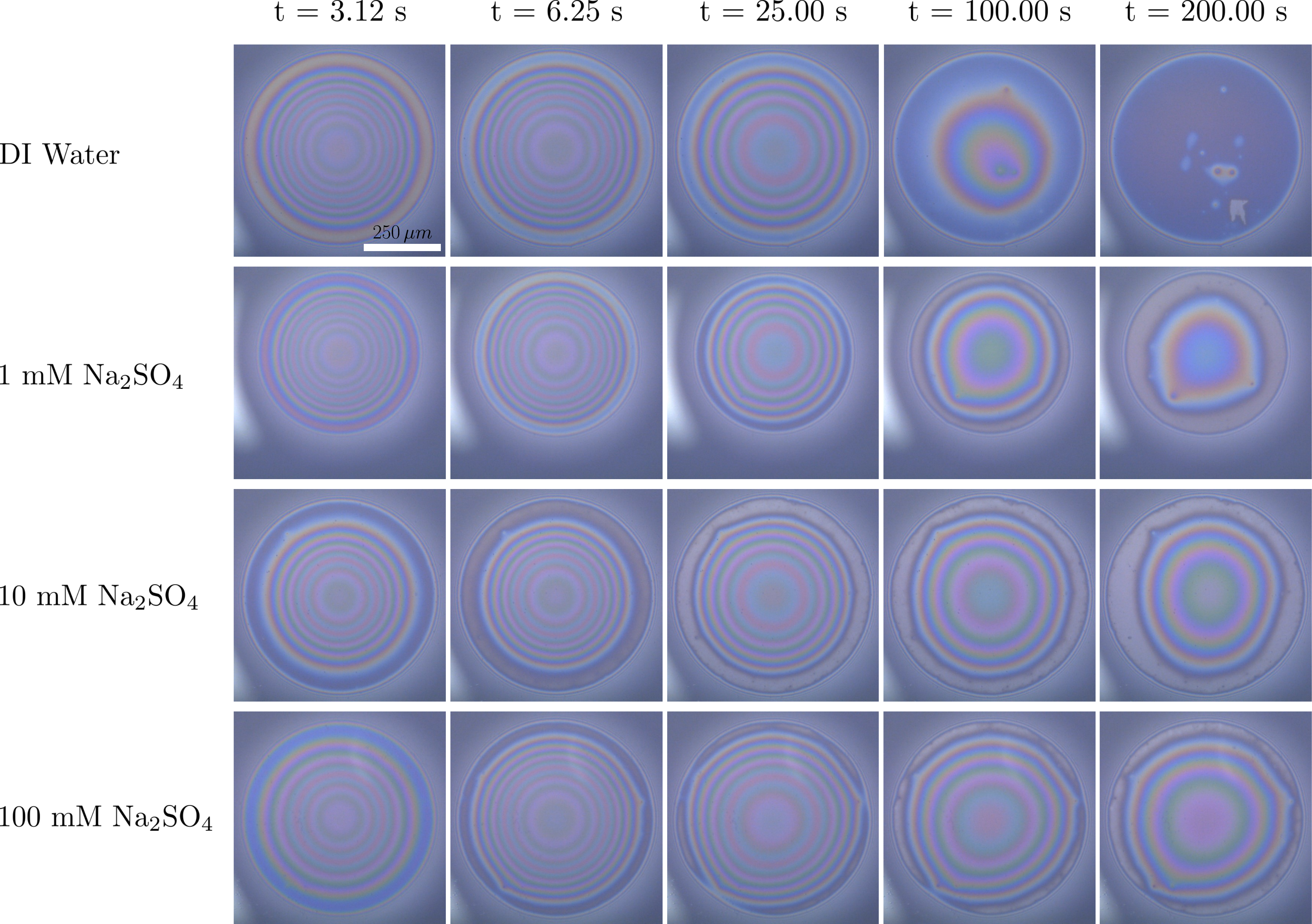}
    \caption{Color interferograms recorded at selected time instants during film drainage for deionised water and Na$_2$SO$_4$ solutions of increasing concentration. Rows correspond to different electrolyte concentrations and columns to successive times after the onset of bubble approach.}
    \label{fig_DIWater_Na2S04_AllConc}
\end{figure}  

Figures~\ref{fig_DIWater_NaCl_AllConc} and~\ref{fig_DIWater_Na2S04_AllConc} present representative interferograms for NaCl and Na$_2$SO$_4$ solutions at different concentrations and time instants, respectively, together with the corresponding results for DI water. At early times, all systems exhibit nearly concentric fringes, indicating an axisymmetric dimple and uniform drainage. As thinning proceeds, the number of fringes decreases, and the fringes broaden radially, consistent with progressive dimple flattening.

With increasing Na$_2$SO$_4$ concentration, a pronounced retardation of film drainage is observed, as evidenced by the persistence of a greater number of interference fringes at comparable times relative to NaCl solutions. This contrast is particularly evident at 10~mM concentration when comparing NaCl (Figure~S3) with Na$_2$SO$_4$ (Figure~\ref{fig_DIWater_Na2S04_AllConc}). In contrast to NaCl, the Na$_2$SO$_4$ interferograms retain a higher degree of radial symmetry over longer times, with comparatively weaker angular non-uniformities in fringe spacing. This indicates that divalent sulfate ions retard drainage while preserving a more spatially uniform film morphology.


\section{Evolution of film characteristics for individual NaCl solutions}\label{Appdx_NaCl_AllPara}

    \begin{figure}
\centering
\includegraphics[width=0.9\linewidth]{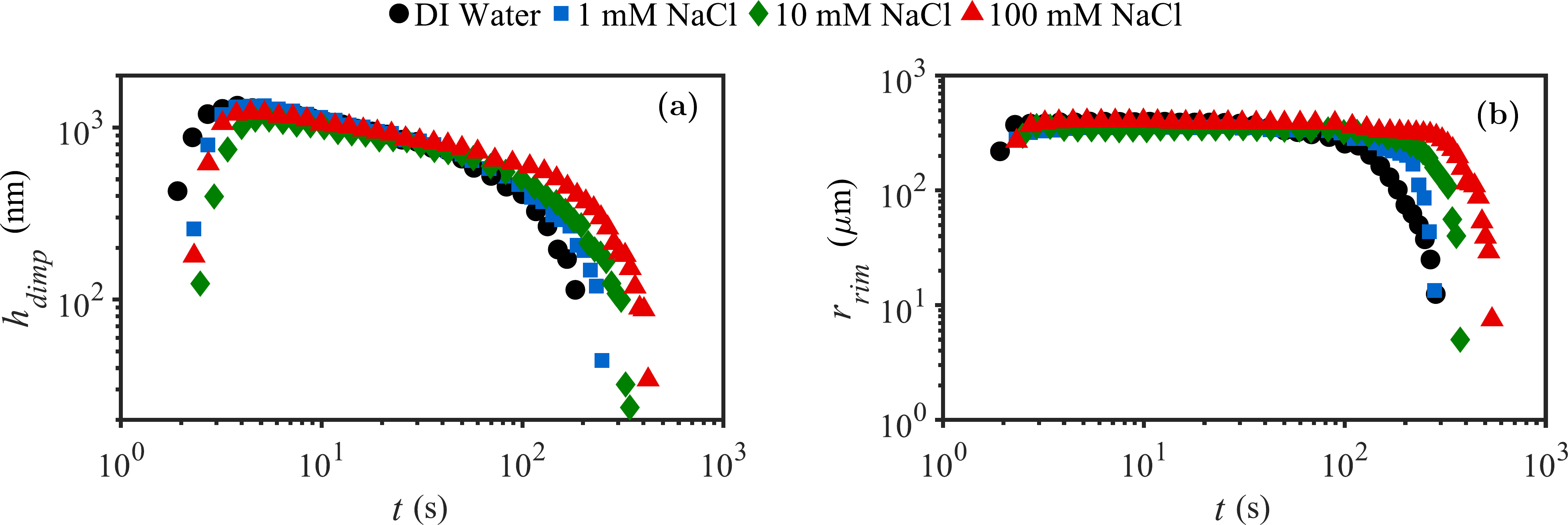}
\caption{Temporal evolution of key geometric descriptors of the draining thin film for DI water and NaCl solutions at concentrations of 1, 10, and 100~mM. (a) Dimple height, $h_{\mathrm{dimp}}$, and (b) rim radius, $r_{\mathrm{rim}}$, are shown as functions of time.}
\label{fig_h_dimpNr_rim}
\end{figure}

Figure \ref{fig_h_dimpNr_rim} shows the temporal evolution of the dimple height, $h_{\mathrm{dimp}}$, and rim radius, $r_{\mathrm{rim}}$, for DI water and NaCl solutions at different concentrations. For the baseline case of DI water, $r_{\mathrm{rim}}$ initially increases during the early stage of drainage, reflecting the lateral expansion of the dimple, and subsequently remains nearly constant over an intermediate time window. At later times, $r_{\mathrm{rim}}$ gradually decreases as the dimple relaxes and the film approaches its stabilized configuration. In parallel, $h_{\mathrm{dimp}}$ exhibits a monotonic decrease following its initial formation, closely mirroring the evolution of the central film thickness. 

With an increase in NaCl concentration, $h_{\min}$ remains significantly smaller than $h_{\max}$ during most of the drainage process and the temporal profile of the dimple height $h_{\mathrm{dimp}}$ closely follows that of $h_{\max}$ (Figure~2(c)). The corresponding rim radius $r_{\mathrm{rim}}$, shown in Figure~\ref{fig_h_dimpNr_rim}(b), initially increases and reaches a maximum values of $359.15$, $375.54$, and $395.56~\mu$m at $3.19$, $2.92$, and $2.73$~s for the 1, 10, and 100~mM solutions, respectively. This early growth reflects the radial spreading of liquid driven by the hydrodynamic pressure generated as the bubble approaches the glass substrate. Notably, the time at which $r_{\mathrm{rim}}$ attains its maximum does not coincide with the instant of maximum dimple height $h_{\max}$, indicating that radial spreading of the rim saturates before the vertical deformation of the dimple reaches its peak. Following this stage, $r_{\mathrm{rim}}$ remains nearly steady until the end of the first drainage regime (65.27~s, 66.25~s, and 88.44~s for 1, 10, and 100mM, respectively), beyond which it decreases as the dimple gradually flattens.

\section{Evolution of film characteristics for Na$_2$SO$_4$ solutions}\label{Appdx_Na2SO4_AllPara}

\begin{figure}
\centering
\includegraphics[width=\linewidth]{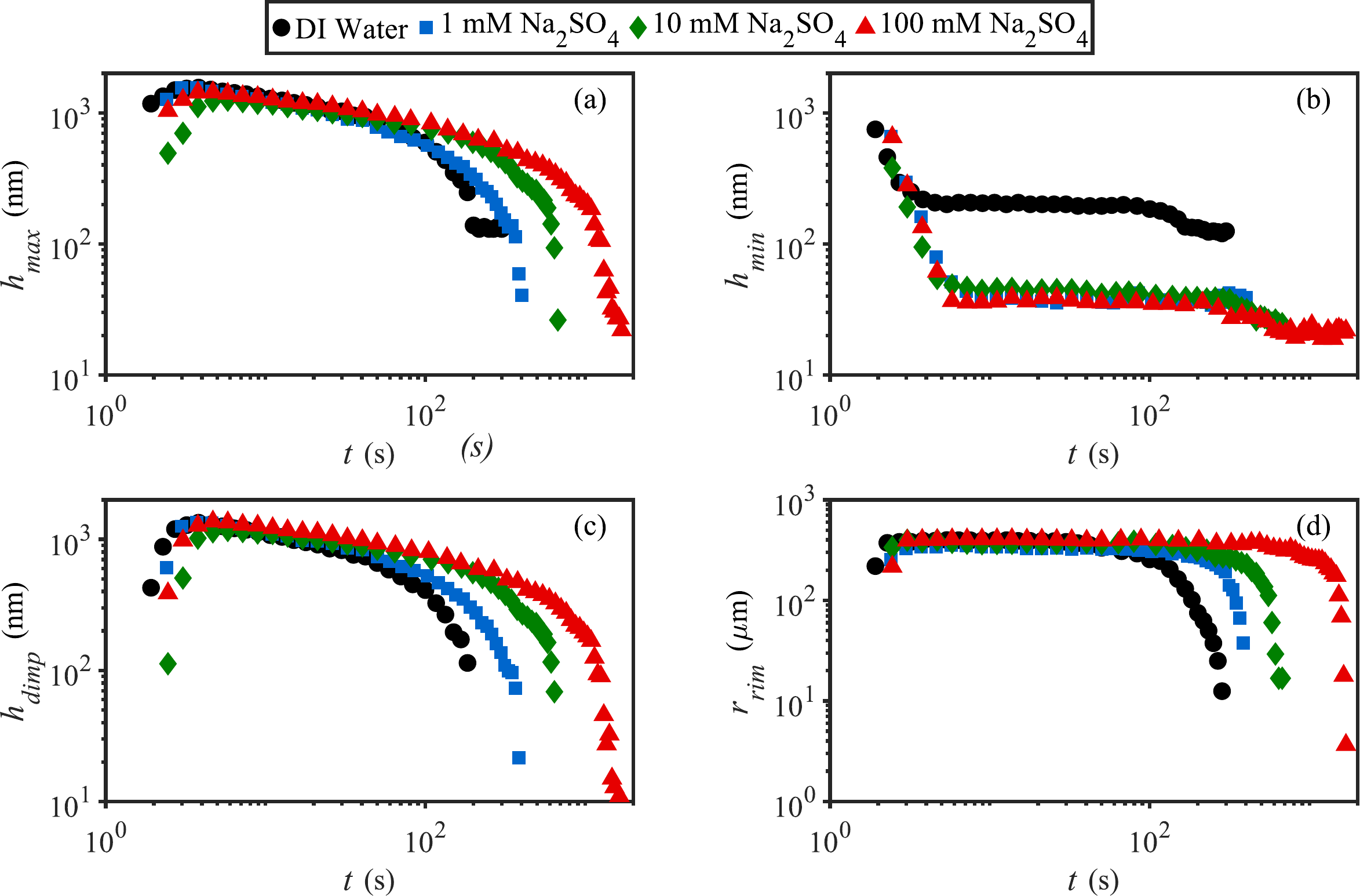}
\caption{Comparison of the temporal evolution of the characteristic geometric parameters for DI water and Na$_2$SO$_4$ solutions of varying concentrations ($1$, $10$, and $100~\mathrm{mM}$). The subplots display the time-dependent variation of: (a) maximum film thickness of the dimple, $h_{\max}$, (b) minimum film thickness at the rim, $h_{\min}$, (c) dimple height, $h_{\mathrm{dimp}}$, and (d) the radius of the rim, $r_{\mathrm{rim}}$.}
\label{fig_CompAllCharAllSol_Na2SO4}
\end{figure}

To quantify the influence of ion valency on the drainage dynamics, we analyze the temporal evolution of the characteristic film parameters, $h_{\max}$, $h_{\min}$, $h_{\mathrm{dimp}}$, and $r_{\mathrm{rim}}$, as shown in Figure~\ref{fig_CompAllCharAllSol_Na2SO4}. While the overall drainage dynamics qualitatively resemble those observed for NaCl, the magnitude of the hydrodynamic retardation and interfacial deformation is significantly amplified, revealing critical ion-specific distinctions.

As illustrated in Figure~\ref{fig_CompAllCharAllSol_Na2SO4}(a), the maximum film thickness $h_{\max}$ for Na$_2$SO$_4$ solutions decays significantly more slowly than in DI water, with the retardation becoming progressively stronger at higher concentrations. The central thickness initially increases, reaching peak values of $1526$, $1430$, and $1449$~nm at approximately $3.07$, $3.77$, and $3.94$~s for the $1$, $10$, and $100$~mM concentrations, respectively. Following these peaks, the decay of $h_{\max}$ proceeds faster, with the first drainage regime persisting for $85.10$~s, $197.04$~s, and $323.14$~s for the three respective concentrations. The subsequent approach toward equilibrium is also significantly prolonged, with $h_{\max}$ stabilizing only after $401$~s, $660$~s, and $1680$~s, respectively (Table~1).

The evolution of the rim thickness $h_{\min}$ (Figure~\ref{fig_CompAllCharAllSol_Na2SO4}(b)) follows the expected trend of decreasing equilibrium thickness with increasing ionic strength, approaching stabilized values of $40$, $26$, and $21$~nm for the $1$, $10$, and $100$~mM Na$_2$SO$_4$ solutions, respectively. However, a more revealing indicator of the drainage dynamics is the time required for $h_{\min}$ to reach these stabilized values. In NaCl solutions, stabilization occurs relatively rapidly (approximately $6.06$~s, $6.41$~s, and $129$~s for $1$, $10$, and $100$~mM). In contrast, the corresponding stabilization times for Na$_2$SO$_4$ are significantly prolonged, occurring at approximately $9.05$~s, $457$~s, and $741$~s.

Because $h_{\min}$ remains much smaller than $h_{\max}$ throughout this extended drainage process, the dimple height $h_{\mathrm{dimp}}$ (Figure~\ref{fig_CompAllCharAllSol_Na2SO4}(c)) closely mirrors the evolution of $h_{\max}$. Consequently, the central dimple remains large and persistent, particularly at $100$~mM, effectively trapping liquid beneath the interface for extended durations.

A further signature of this sulfate-induced retardation appears in the lateral rim radius $r_{\mathrm{rim}}$ (Figure~\ref{fig_CompAllCharAllSol_Na2SO4}(d)). For $1$, $10$, and $100$~mM Na$_2$SO$_4$, $r_{\mathrm{rim}}$ reaches its initial maximum at approximately $3.07$, $3.77$, and $3.82$~s, attaining peak values of $354.58$, $393.54$, and $400.74~\mu$m, respectively. It then remains nearly constant throughout the prolonged first drainage regime. Beyond this stage, all concentrations, most prominently $100$~mM exhibit a clear contraction of $r_{\mathrm{rim}}$ as the film transitions from a broad, extended dimple to a highly confined, lenticular geometry.

This high degree of spatial confinement at the rim and retardation in film drainage is a direct consequence of the specific effects of ion valency on the electrical double layer (EDL). Because Na$_2$SO$_4$ dissociates into two Na$^+$ ions and one divalent SO$_4^{2-}$ ion, it yields a threefold higher ionic strength compared to NaCl at the same molarity. This increase in ionic strength significantly reduces the Debye length ($\lambda_D \propto 1/\sqrt{I}$) by approximately $30$--$40\%$ relative to NaCl (cf. Tables~1). Consequently, the repulsive disjoining pressure at the rim is screened much more effectively, allowing the local film thickness to collapse to smaller values than those observed in monovalent solutions. For instance, at $t = 10$~s, the rim thickness for $1$~mM Na$_2$SO$_4$ is approximately $40$~nm, compared to about $56$~nm for $1$~mM NaCl.

If electrostatic screening alone governed the drainage, the shorter Debye length in the Na$_2$SO$_4$ compared to NaCl system would be expected to promote faster thinning of the film once the film reaches the nanometric regime. Instead, the markedly extended stabilization times and the pronounced contraction of the rim demonstrate that an additional retarding mechanism also influence the drainage dynamics. In the present system, this mechanism arises from enhanced Marangoni stresses associated with the distinct interfacial thermodynamics of the divalent sulfate anion, SO$_4^{2-}$.

Being a strongly hydrated kosmotrope (hydration number $\approx 12$) \cite{chizhik2002microstructure, marcus2009effect}, the SO$_4^{2-}$ ion is repelled from the air--water interface much more vigorously than the monovalent chloride ion, Cl$^-$ \citep{gopalakrishnan2005air,seki2023ions}. Physically, this enhanced exclusion arises from image-charge forces that scale with the square of the ion valency \cite{onsager1934surface}, imposing an electrostatic penalty on the divalent sulfate ion ($z=2$) that is approximately four times greater than that on the chloride ion. As the film thins, particularly at the rim where the area-to-volume ratio is highest, this severe negative adsorption forces a larger fraction of SO$_4^{2-}$ ions to desorb into the adjacent finite bulk layer.

This interfacial exclusion quantitatively manifests in a higher solutal surface activity, $d\sigma/dc$, which is approximately $1.5$ times larger for Na$_2$SO$_4$ ($\approx 2.73~\mathrm{mN\, m^{-1}\,M^{-1}}$) than for NaCl ($\approx 1.79~\mathrm{mN\,m^{-1}\,M^{-1}}$) \citep{weissenborn1996surface}. Since the Marangoni stress scales as $\Delta\sigma \propto c (d\sigma/dc)^2$, the resulting retardation is strongly amplified. For instance, for $100$~mM Na$_2$SO$_4$, the theoretical surface tension increase at the thinned rim ($h \approx 30$~nm) rises to $\Delta \sigma \approx 0.0100~\mathrm{mN\,m^{-1}}$, more than double the value calculated for NaCl ($\Delta \sigma \approx 0.0043~\mathrm{mN\,m^{-1}}$). This disparity is rooted in the significantly larger negative surface excess of the sulfate system ($\Gamma \approx -1.1 \times 10^{-7}~\mathrm{mol\,m^{-2}}$ compared to $-7.1 \times 10^{-8}~\mathrm{mol\,m^{-2}}$ for NaCl). Consequently, the combination of stronger ion exclusion and the quadratic sensitivity of surface tension to surface activity, $d\sigma/dc$, generates an inward Marangoni flow that opposes the outward hydrodynamic  drainage (see Figure~3(a)),
resulting in the pronounced retardation of drainage compared to that in NaCl solution.

Overall, these geometric signatures and physical mechanisms demonstrate that the Na$_2$SO$_4$ system operates in a regime of strong confinement and pronounced drainage retardation. Enhanced electrostatic screening compresses the electrical double layer, allowing the rim to collapse to significantly smaller thicknesses and thereby creating a severe nanoscale hydraulic bottleneck that dominates the drainage resistance. Although Marangoni stresses generated by ion redistribution further oppose interfacial flow and contribute to the retardation of drainage, their effect is secondary to the geometric confinement imposed by the thinned rim. The combined action of these mechanisms traps liquid beneath the dimple, sustaining a deep and steep-walled morphology and producing the slowest drainage rates observed in this study, most prominently for the $100$~mM Na$_2$SO$_4$ solution.

\section{Scaling Analysis}

From experiments, we note that after the initial contact, drainage of the trapped thin film occurs in three distinct regimes. In the first regime (Regime-I), the thickness at the center of the film varies as $h_{max} \propto t^{-1/4}$. In the second regime (Regime-II), we obtain an exponential drainage (based on plots in semi-log), $h_{max} \propto e^{-t/\tau}$ and in the final regime (Regime-III), a steep slope for drainage is observed before an equilibrium uniform film is obtained. 

\subsection{Scaling analysis for Regime-I}

	\begin{figure}
		\centering		
        \includegraphics[width=0.8\linewidth]{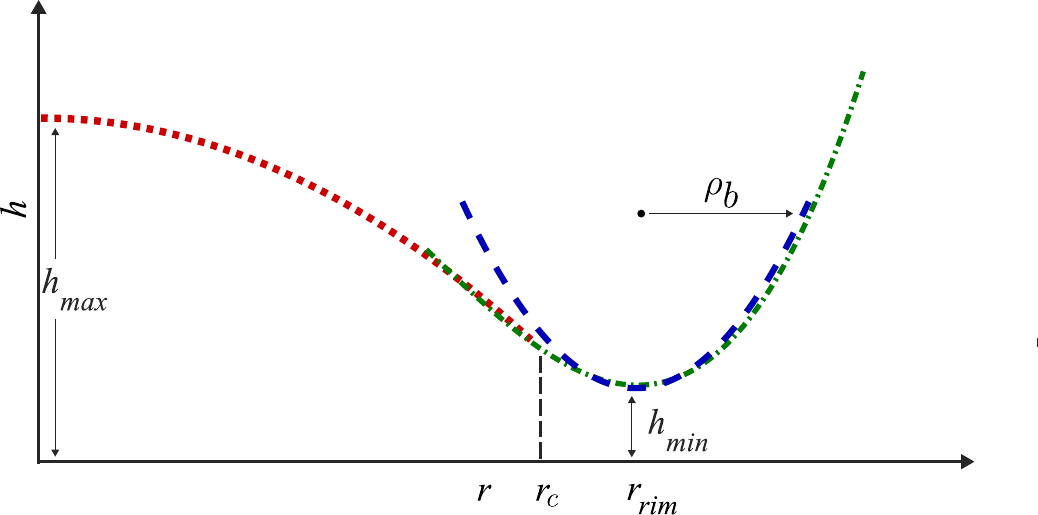}
        \caption{Radial film thickness profile $h(r)$ illustrating the dimple geometry. The central region is approximated by a quadratic profile (red dotted), while the barrier-ring region is represented by a cubic profile (green dashed), showing the minimum $h_{\min}$ at $r = r_{\mathrm{rim}}$. The profiles are smoothly matched. The blue dash-dotted curve indicates the local curvature, and $\rho_b$ represents the radius of curvature in the barrier region.}
    \label{fig_dimpleFit}
	\end{figure}

Here, we provide the complete derivation of the power-law scaling $h_{max}(t) \propto t^{-1/4}$ observed in Regime I in Figure~9 of the manuscript. Motivated by the scaling analysis by Bluteau \textit{et al.} \cite{bluteau2017water} for a dimpled water film sandwiched between an oil droplet and a glass substrate, we derive scalings for bubble-substrate interactions. This regime is characterized by a balance between the lubrication pressure drop and capillary forces, leading to a visco-capillary drainage regime. 

Based on the geometric evolution of the film, we note that the film thickness variation in the dimple region (similar to that for the drop in Bluteau \textit{et al.} \cite{bluteau2017water}) can be approximated as (see Figure~\ref{fig_dimpleFit}):
\begin{equation}
    h_d(r) = h_{max}(t) \left(1 - \left(\frac{r}{r_{rim}}\right)^2 \right)
    \label{eq:parabolafilmdimple}
\end{equation}
where $r_{rim}$ is the extent of the rim region with minimum of the film thickness at $r = r_{rim}$.
By choosing an expansion about $r = r_{rim}$, we can write the film thickness in rim region, $h_{b}(r)$, as:
\begin{equation}
    h_{b}(r) = h_{min} + \frac{(r-r_{rim})^2}{2}h_{rr}|_{r=r_{rim}}+h_{rrr}|_{r = r_{rim}}\frac{(r-r_{rim})^3}{6}+\mathcal{O}((r-r_{rim})^3)
    \label{eq:filmrim}
\end{equation}
where $h_{rr}|_{r=r_{rim}}$ is the curvature of the profile at $r=r_{rim}$.
The two profiles, the one around the dimple and the other around the rim, are matched at $r = r_c$ such that $h_d(r_c) = h_b(r_c)$, 
$h_d'(r_c) = h_b'(r_c)$ and $h_d''(r_c) = h_b''(r_c)$. Hence, we obtain:
\begin{eqnarray}
    r_c &=& r_{rim}\left(1 + \frac{3 h_{min}}{2 h_{max}(t)}\right)\\
    h_{rr}|_{r=r_{rim}} &=& \frac{2 h_{max}(t) (4 h_{max}(t) - 3 h_{min})}{3 h_{min} r_{rim}^2} \approx \frac{8 h_{max}(t)^2}{3 h_{min} r_{rim}^2}\\
    h_{rrr}|_{r=r_{rim}} &=& \frac{16 h_{max}(t)^3}{9 h_{min}^2 r_{rim}^3}
\end{eqnarray}

Assuming validity of the lubrication approximation, radial velocity can be written:
\begin{equation}
    u_r(r,z) = \frac{1}{2\mu}\frac{dp}{dr} z (\beta h(r,t) - z)
    \label{eq:ur}
\end{equation}
where $\beta = 2$ corresponds to shear free interface (fully mobile) and $\beta = 1$ corresponds to no-slip condition at the bubble-liquid interface; the liquid-glass surface is assumed to be no-slip. 
Volumetric flux per unit perimeter thus can be written as:
\begin{equation}
q(r) = -\frac{dp}{dr}\frac{(3\beta-2)h^3}{12\mu}.
\label{eq:qr}
\end{equation}
Further, kinematic condition for the thin film yields:
\begin{equation}
q(r) = -\frac{\int_0^r\left(r\frac{\partial h}{\partial t}\right)dr}{r}.
\label{eq:qrdhdt}
\end{equation}

The leading order term in the lubrication pressure gradient at $r = r_c$, obtained from eqns \ref{eq:parabolafilmdimple}, \ref{eq:filmrim}, \ref{eq:qr}, \ref{eq:qrdhdt}, is:
\begin{equation}
\frac{dp}{dr} = \frac{r_{rim} \mu h_{max}'(t)}{9 (3 \beta -2) h_{min}^3}
\label{eq:dpdr_rc}
\end{equation}
whereas, the pressure drop at $r = r_{rim}$ is given by:
\begin{equation}
\frac{dp}{dr} = \frac{3 r_{rim} \mu h_{max}'(t)}{(3 \beta -2) h_{min}^3}
\label{eq:dpdr_rd}
\end{equation}
which is of the same form as \ref{eq:dpdr_rc} but higher by a factor of 27. Pressure gradient in the bulk of the dimple away from $r = r_c$ is $\propto 1/h_{max}^3$ and is much smaller than $1/h_{min}^3$. Thus, we conclude that most of the pressure drop occurs in the rim region $r \in (r_c,r_{rim})$. Further, performing normal stress balance at the bubble-liquid interface, pressure variation due to curvature and disjoining pressure can be written as:
\begin{equation}
p(r) = P_{ext} + \frac{2 \sigma}{R_L} - \sigma \left(h_{rr} + \frac{h_r}{r}\right) - \Pi_{EDL}(h) - \Pi_{vdW}(h)
\label{eqn:pressure}
\end{equation}
where $\Pi_{EDL} = 64 k_B T n_{\infty} tanh^2[(z_i e\psi_0)/(4 k_B T)] e^{-h/\lambda_D}$ (valid for $h >> \lambda_D$)
and $\Pi_{vdW} = A_{eff}/h^3$. Here, $n_\infty$ is the concentration of ions in the bulk, $\lambda_D$ is the Debye-H\"{u}ckel length scale, $k_B$ is the Boltzmann constant, $T$ is the temperature, $z_i$ is the valency of the ions and $e$ is the electron charge. In the expression for $\Pi_{vdW}$, $A_{eff}$ is the effective Hamaker constant for the thin film of the electrolyte solution on a glass substrate. In the current regime (film thickness $\sim 1\mu$m - 400nm??), the drainage is primarily driven by capillary forces. Thus, neglecting the disjoining pressure terms and comparing the gradient in pressure due to capillary forces and the viscous forces in the rim region, we get:
\begin{equation}
    h_{min} = -\frac{27 r_{rim}^4 \mu h_{max}'(t)}{16 (3\beta -2)\sigma h_{max}(t)^3}.
\label{eq:hmin}
\end{equation}
Further, equating the pressure difference $p(0)-P_{ext}$ with the net pressure drop in the film due to viscous terms, we get:
\begin{equation}
\frac{2\sigma}{R_L} + \frac{4 \sigma h_{max}(t)}{r_{rim}^2} =  \int_0^\infty \frac{12 q(r) \mu}{(3 \beta - 2) h(r)^3}dr.
\end{equation}
As discussed earlier, the pressure drop is essentially dominated by pressure drop in the rim region, thus, we can write:
\begin{equation}
    \frac{2 \sigma}{R_L} +  \frac{4 \sigma h_{max}(t)}{r_{rim}^2} =  \int_{r_c}^{r_{rim}} \frac{12 q(r_{rim}) \mu}{(3 \beta - 2) h_{min}^3}dr = \frac{12 q(r_{rim}) \mu}{(3 \beta - 2) h_{min}^3}(r_{rim} - r_c)
    \label{eq:dhdtRegimeI}
\end{equation}
under the approximations $q(r_c) \approx q(r_{rim})$ and $h(r) \sim h_{min}$ in the rim region around $r = r_{rim}$, where the film is essentially flat. In the leading order, $q(r_{rim}) \sim -h_{max}'(t) r_{rim}/4$. Here, $r_c$ is the radial location for matching the dimple and rim profiles:
\begin{equation}
r_{rim} - r_c =  \frac{3 h_{min} r_{rim}}{2 h_{max}(t)}.
\label{eq:exp-rc}
\end{equation}.
Further, we note that the ratio of the second and first terms in LHS of the above equation is given by $2 h_{max}(t) R_L/r_{rim}^2 <<1$.  Thus, substituting $h_{min}$ from Eq. \ref{eq:hmin}, we can write the Eq. \ref{eq:dhdtRegimeI} as:
\begin{equation}
  \frac{1}{h_{max}^5}\frac{dh_{max}}{dt} =  -\frac{64 (3 \beta - 2)}{81}\frac{\sigma R_L}{r_{rim}^6 \mu}
\end{equation}
Thus, we get:
\begin{equation}
h_{max}(t) = \frac{3}{4}\left(\frac{\mu r_{rim}^6} {(3 \beta-2)\sigma R_L}\right)^{1/4} t^{-1/4}
\end{equation}
For free-shear surface $\beta = 2$, we get:
\begin{equation}
h_{max}(t) = \frac{3}{16}\left(\frac{\mu r_{rim}^6} { \sigma R_L}\right)^{1/4} t^{-1/4},
\end{equation}
whereas, for immobile surface ($\beta = 1$):
\begin{equation}
h_{max}(t) = \frac{3}{4}\left(\frac{\mu r_{rim}^6} {\sigma R_L}\right)^{1/4} t^{-1/4}.
\end{equation}
We note that the $t^{-1/4}$ power law does not depend on the nature of the boundary condition at the bubble-liquid interface.

\subsection{Scaling analysis for Regime II}
In this regime, the rim settles to an equilibrium thickness given by the balance between the capillary pressure $2 \sigma/R_L$ and $\Pi_{EDL}$. Thus, we get the minimum thickness in the rim region as:
\begin{equation}
    h_{min} = h_{eq} = - 2 \lambda~ln\left(\frac{2 \sigma}{K R_L}\right)
\end{equation}
where $K = 64 k_B T n_{\infty} tanh^2[(z_i e\psi_0)/(4 k_B T)]$.
Substituting, $h_{min} = h_{eq}$ in Eq.\ref{eq:dhdtRegimeI}, we obtain:
\begin{equation}
   \frac{2\sigma}{R_L} \sim -\frac{18 r_{rim}^2 h_{max}'(t)\mu}{4(3 \beta - 2)h_{eq}^2 h_{max}(t)}
\end{equation}
The dimple thickness obtained as:
\begin{equation}
    h_{max}(t) \propto e^{-t/\tau}
\end{equation}
where the relaxation time is given by:
\begin{equation}
    \tau = \frac{9  R_L\mu r_{rim}^2}{4(3\beta-2) \sigma h_{eq}^2 }.
\end{equation}

As the film relaxes towards flattening of the dimple region due to drainage, overall film thickness reduces.
Finally, Regime-III is given by drainage of an essentially thin film where $h_{max}(t) \sim h_{min} \sim h_{eq}$ where in the entire film, including the dimple, disjoining pressure cannot be neglected.

\subsection{Scaling analysis for Regime-III}

As the film relaxes and the dimple flattens, the drainage enters Regime-III, characterized by an essentially uniform, thin film where the disjoining pressure can no longer be neglected. In this late stage, we model the system assuming a fully mobile (zero-shear) boundary condition at the bubble-liquid interface.

Assuming the lubrication approximation for a thin uniform film of thickness $h$ draining between a solid substrate and a fully mobile bubble interface, we obtain the pressure gradient in the film as,
	\begin{equation}
		\frac{dp}{dr} = \frac{12\mu r}{2(3\beta -2)h^3} \frac{dh}{dt}
	\end{equation}
Integrating this from the center ($r = 0$) to the radial extent of the contact zone ($r = R_c$), where the pressure reaches the bulk atmospheric value, we obtain the total viscous pressure drop:
\begin{equation}
    \Delta p = -\frac{12\mu R_c^2}{4(3\beta -2)h^3} \frac{dh}{dt}
\end{equation}

The rate of film thinning is determined by the balance between the driving capillary pressure ($2\sigma/R_L$), the opposing repulsive electrical double layer (EDL) disjoining pressure ($\Pi_{EDL} = K e^{-h/\lambda}$), and this viscous hydrodynamic resistance. Equating these yields the governing force balance for Regime-III:
\begin{equation}
    \frac{2\sigma}{R_L} - K e^{-h/\lambda} = -\frac{12\mu R_c^2}{4(3\beta-2)h^3} \frac{dh}{dt}
\end{equation}
where $K = 64 k_B T n_{\infty} \tanh^2[(z_i e\psi_0)/(4 k_B T)]$.

\subsubsection*{Equilibrium State and Perturbation Analysis}

Once the macroscopic dimple relaxes to a nearly uniform, planar geometry ($h ~ h_{\mathrm{eq}}$), drainage slows considerably. The capillary pressure is perfectly balanced by the disjoining pressure:
\begin{equation}
    \frac{2\sigma}{R_L} = K e^{-h_{\mathrm{eq}}/\lambda} \implies h_{\mathrm{eq}} = \lambda \ln\left( \frac{R_L K}{2\sigma} \right)
\end{equation}

To analyze the late-stage drainage dynamics approaching this limit, we assume the film thickness $h(t)$ is extremely close to the equilibrium thickness. We define a perturbation variable, $\tilde{h}(t)$, representing this tiny difference:
\begin{equation}
    h(t) = h_{\mathrm{eq}} + \tilde{h}(t)
\end{equation}
Because we are in the final stages of thinning, $\tilde{h}(t) \ll h_{\mathrm{eq}}$. Since $h_{\mathrm{eq}}$ is a constant, the temporal derivative is simply $dh/dt = d\tilde{h}(t)/dt$. Substituting $h = h_{\mathrm{eq}} + \tilde{h}(t)$ back into the original force balance gives:
\begin{equation}
    \frac{2\sigma}{R_L} - K e^{-(h_{\mathrm{eq}} + \tilde{h}(t))/\lambda} = -\frac{12\mu R_c^2}{4(3\beta-2)(h_{\mathrm{eq}} + \tilde{h}(t))^3} \frac{dH}{dt}
\end{equation}
Or,
\begin{equation}
    \frac{2\sigma}{R_L} - K e^{-h_{\mathrm{eq}}/\lambda} \left[ e^{-\tilde{h}(t)/\lambda} \right] = -\frac{12\mu R_c^2}{4(3\beta-2)h_{\mathrm{eq}}^3 \left(1 + \frac{\tilde{h}(t)}{h_{\mathrm{eq}}}\right)^3} \frac{d\tilde{h}(t)}{dt}
\end{equation}
Recognizing that $K e^{-h_{\mathrm{eq}}/\lambda} = \frac{2\sigma}{R_L}$, we substitute this into the left side and factor out the capillary pressure term:
\begin{equation}
    \frac{2\sigma}{R_L} \left( 1 - e^{-H/\lambda} \right) = -\frac{12\mu R_c^2}{4(3\beta-2)h_{\mathrm{eq}}^3 \left(1 + \frac{H}{h_{\mathrm{eq}}}\right)^3} \frac{dH}{dt}
\end{equation}
Because the perturbation $\tilde{h}(t)$ is extremely small ($\tilde{h}(t) \ll h_{\mathrm{eq}}$ and $\tilde{h}(t) \ll \lambda$), we apply Taylor series approximations to linearize the equation. On the right side, $\tilde{h}(t)/h_{\mathrm{eq}} \approx 0$. On the left side, the exponential decay is linearized as $e^{-\tilde{h}(t)/\lambda} \approx 1 - \tilde{h}(t)/\lambda$. Applying these approximations reduces the non-linear equation to:
\begin{equation}
    \frac{2\sigma}{R_L} \left( \frac{\tilde{h}(t)}{\lambda} \right) = -\frac{12\mu R_c^2}{4(3\beta-2)h_{\mathrm{eq}}^3} \frac{d\tilde{h}(t)}{dt}
\end{equation}
Rearranging the equation to isolate the derivative $dH/dt$:
\begin{equation}
    \frac{d\tilde{h}(t)}{dt} = - \left( \frac{8(3\beta-2)\sigma h_{\mathrm{eq}}^3}{12\mu R_c^2 R_L \lambda} \right) \tilde{h}(t)
\end{equation}
By inspecting the grouped constants, we can explicitly define the characteristic relaxation time for Regime-III:
\begin{equation}
    \tau_2 = \frac{12\mu R_c^2 R_L \lambda}{8(3\beta-2)\sigma h_{\mathrm{eq}}^3}
\end{equation}
Integrating the differential equation yields $\ln(\tilde{h}(t)) = -t/\tau_2 + C$. Taking the exponential of both sides and substituting $\tilde{h}(t)(t) = h(t) - h_{\mathrm{eq}}$ gives the final mathematical proof for Regime-III exponential decay under fully mobile boundary conditions:
\begin{equation}
    h(t) = h_{\mathrm{eq}} + \tilde{h}_0(t) \exp\left( - \frac{t}{\tau_2} \right)
\end{equation}
where $\tilde{h}_0(t)$ is the initial perturbation from equilibrium at the onset of Regime-III.

\section{Universal Scaling and Comparison with Literature}

\begin{figure}
\centering
\includegraphics[width=1\linewidth]{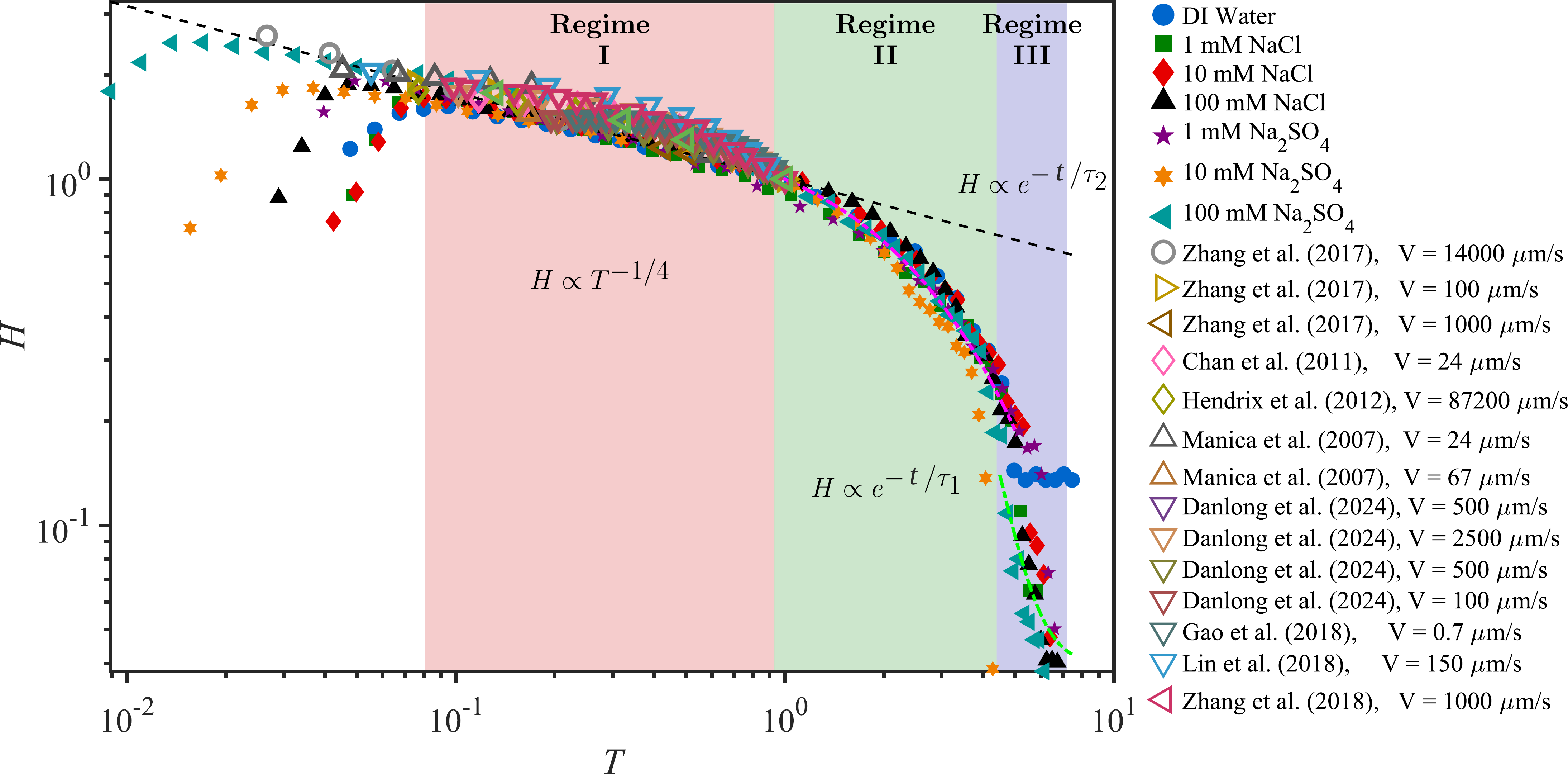}
\caption{Comparison of the dimensionless film thickness $H = h_{\max}/h_0$ as a function of dimensionless time $T = t/t_0$ with previously reported experimental data from the literature. Data from the present study for DI water, NaCl, and Na$_2$SO$_4$ are shown alongside literature results spanning a wide range of systems and experimental conditions. The early-time dynamics collapse onto the universal visco-capillary scaling $H \propto T^{-1/4}$ (Regime~I), demonstrating excellent agreement with prior studies.}
\label{fig_scaling_literature}
\end{figure}

Figure~\ref{fig_scaling_literature} compares the dimensionless drainage dynamics obtained in the present study with a broad range of previously reported experimental data spanning different fluids, substrates, and operating conditions. Despite substantial variations in approach velocity, fluid properties, and interfacial conditions, the early-time dynamics exhibit a remarkable collapse onto the universal visco-capillary scaling $H \propto T^{-1/4}$. This agreement indicates that the initial drainage is governed by a robust balance between viscous resistance and capillary pressure associated with the dimpled film geometry, independent of electrolyte composition or interfacial physicochemistry. The consistency across disparate datasets confirms that Regime~I represents a universal hydrodynamic regime dictated primarily by geometry and bulk fluid properties. Deviations from this scaling at later times arise from electrolyte-specific effects, including Marangoni stresses and disjoining-pressure-driven confinement, which progressively alter the drainage dynamics beyond the visco-capillary limit.


\subsubsection*{Analytical Derivation of Transition Scales}

The characteristic transition scales $(t_0, h_0)$ are analytically derived by matching the variation of $h_{max}$ in both Regime-I and Regime-II.

Regime~I is given by,  $h_{\max}(t) = C_1 t^{-1/4}$ (where $C_1 = \frac{3}{4} [ \mu r_{\mathrm{rim}}^6 / (\sigma R_L) ]^{1/4}$) and Regime~II is $h_{\max}(t) = h_{0,theory} \exp[-(t - t_{0,theory})/\tau_1]$. By imposing the continuity of both the film thickness, ($h_{\max}$) and its rate of change, ($dh_{\max}/dt$) at $t = t_{0,theory}$, we obtain,
\begin{equation}
    -\frac{1}{4} C_1 t_{0,theory}^{-5/4} = -\frac{h_{0,theory}}{\tau_1}.
\end{equation}
Substituting the continuity condition ($h_{0,theory} = C_1 t_{0,theory}^{-1/4}$) into the above expression gives,
\begin{equation}
    t_{0,theory} = \frac{\tau_1}{4}.
\end{equation}
This establishes that the transition time is universally governed by, and directly proportional to, the relaxation time, $\tau_1$. Substituting $t_{0,theory}$ back into the visco-capillary scaling law and simplifying gives the analytical transition thickness:
\begin{equation}
    h_{0,theory} = \sqrt{\frac{3}{2}} r_{\mathrm{rim}} \sqrt{\frac{h_{\mathrm{eq}}}{R_L}}.
\end{equation}

The transition from Regime~I to Regime~II is triggered by the attainment of the equilibrium thickness at the rim, $h_{\min} = h_{\mathrm{eq}}$, where disjoining-pressure effects become significant. Since the hydrodynamic resistance scales as $h_{\min}^{-3}$, the drainage rate is primarily controlled by the rim thickness. Consequently, the onset of Regime~II is governed by a local condition at the rim rather than the global geometry of the film.

\section{Estimation of measurement uncertainties}

The uncertainties associated with the experimentally measured quantities arise from two independent sources: (i) the positioning accuracy of the microcontroller-driven $z$-stage used to control the bubble motion, and (ii) the optical reconstruction of the thin film thickness using color interferometry.

\subsection{Uncertainty in imposed displacement.}
The bubble is translated toward the substrate over a displacement $\Delta h$ using a motorized $z$-stage (Holmarc MVTS-75-25), whose absolute positioning accuracy is specified as $\pm 5~\mu$m. In addition, the determination of the bubble apex location depends on the measured bubble radius, which carries an uncertainty of approximately $\pm 0.1~\mathrm{mm}$ ($\pm 100~\mu$m). Since these two uncertainties arise from independent measurements, the uncertainty in the imposed displacement is estimated by combining them in quadrature,
\begin{equation}
\delta(\Delta h) = \sqrt{(\delta R_b)^2 + (\delta z)^2},
\end{equation}
where $\delta R_b = 100~\mu$m and $\delta z = 5~\mu$m. This yields
\begin{equation}
\delta(\Delta h) \approx 100~\mu\mathrm{m}.
\end{equation}
Thus, the imposed displacement used in the experiments is
\begin{equation}
\Delta h = 751.5 \pm 100~\mu\mathrm{m}.
\end{equation}

It is important to emphasize that this uncertainty reflects only the accuracy with which the macroscopic bubble position relative to the substrate is determined. It does not influence the measurement of the thin film thickness itself, which is obtained independently from interferometric analysis.

\subsection{Uncertainty in bubble approach velocity.}
The bubble is translated toward the substrate at a programmed constant velocity
\begin{equation}
V_b = 350 \pm 2~\mu\mathrm{m\,s^{-1}},
\end{equation}
where the uncertainty reflects the positioning accuracy of the stage over the imposed displacement.

\paragraph{Uncertainty in translation time.}
The time required for the imposed translation is estimated as
\begin{equation}
t = \frac{\Delta h}{V_b}.
\end{equation}
Propagation of uncertainties gives
\begin{equation}
\left(\frac{\delta t}{t}\right)^2 =
\left(\frac{\delta (\Delta h)}{\Delta h}\right)^2 +
\left(\frac{\delta V_b}{V_b}\right)^2 .
\end{equation}
Substituting the experimental values yields
\begin{equation}
t = 2.15 \pm 0.29~\mathrm{s}.
\end{equation}

\section{Drainage dynamics for slip and no-slip conditions for different thin-film profiles}
Assuming that the thin film shape remains the same during drainage, we write, 
$h(r,t) = h_{\min} + \bigl(h_{\max}(t)-h_{\min}\bigr)f(r)$, where $f(r)$ captures the radial variation in the thin film thickness, with $h_{max}$ as the film thickness at $r = 0$, and $h_{min}$ is the film thickness at the rim.
Then, $\dfrac{\partial h}{\partial t} = \dfrac{d h_{\max}}{dt}\,f(r)$.
For a flat film, we set: $h_{\min}=h_{max} = h(t)$ and $f(r)=1$.

We note that we can modify the expression in Eq.\ref{eq:uavg} for these profiles using,
\begin{equation}
    u_{\text{avg}}(r)
    = -\frac{1}{2\pi r h}\frac{dh_{\max}}{dt}
       \int_0^r 2\pi r f(r) dr
\end{equation}
and write the pressure gradient as,
\begin{equation}
    \frac{\partial p}{\partial r}
    = \frac{12\mu}{(3\beta-2)\,h^3}\cdot\frac{1}{2\pi r h}\cdot
      \frac{d h_{\max}}{dt}\int_0^r (2\pi r\,f(r))\,dr
      \label{eq:dpdr}
\end{equation}

Integrating Eq.\ref{eq:dpdr} and imposing the boundary condition, $p(R_c) = 0$, we get:
\begin{equation}
    p(R_c) - p(r)
    = \left(\frac{d h_{\max}}{dt}\right)
      \int_r^{R_c}
      \left[\frac{12\mu}{(3\beta-2)\,h^3}\cdot\frac{1}{2\pi r}
            \int_0^r 2\pi r\,f(r)\,dr\right]dr
\end{equation}
The total lubrication force balancing the drainage driving force (forces on the bubble) is given by:
\begin{equation}
    F
    = \int_0^{R_c} p(r)\cdot 2\pi r\,dr
    = -\frac{d h_{\max}}{dt}
      \int_0^{R_c} 2\pi r
      \left[\int_r^{R_c}
            \frac{12\mu}{(3\beta-2)}\cdot\frac{1}{2\pi r}\cdot\frac{1}{h^3}
            \int_0^r 2\pi r\,f(r)\,dr\;dr\right]dr
            \label{eq:dhdtEqn}
\end{equation}
which can be written in the following form:
\begin{equation}
      \left(\frac{d h_{\max}}{dt}\right)\cdot
      \left(\frac{24\pi\mu}{3\beta-2}\right)
      \int_0^{R_c}\!\!\int_r^{R_c}
      r\,\frac{1}{r}\cdot\frac{1}{h^3}
      \left(\int_0^r r\,f(r)\,dr\right)dr\,dr + F = 0
\end{equation}

\noindent
The corresponding $h_{\text{avg}}$ is given by,
\begin{equation}
h_{avg} = (1-\alpha) h_{min} + \alpha h_{max}
\label{eq:havg}
\end{equation}
where,
\begin{equation}
    \alpha = \int_0^{R_c} f(r)\cdot {2\pi r\,dr}{\pi R_c^2}
\end{equation}

\begin{figure}[!h]
    \centering
\includegraphics[width=4.5in]{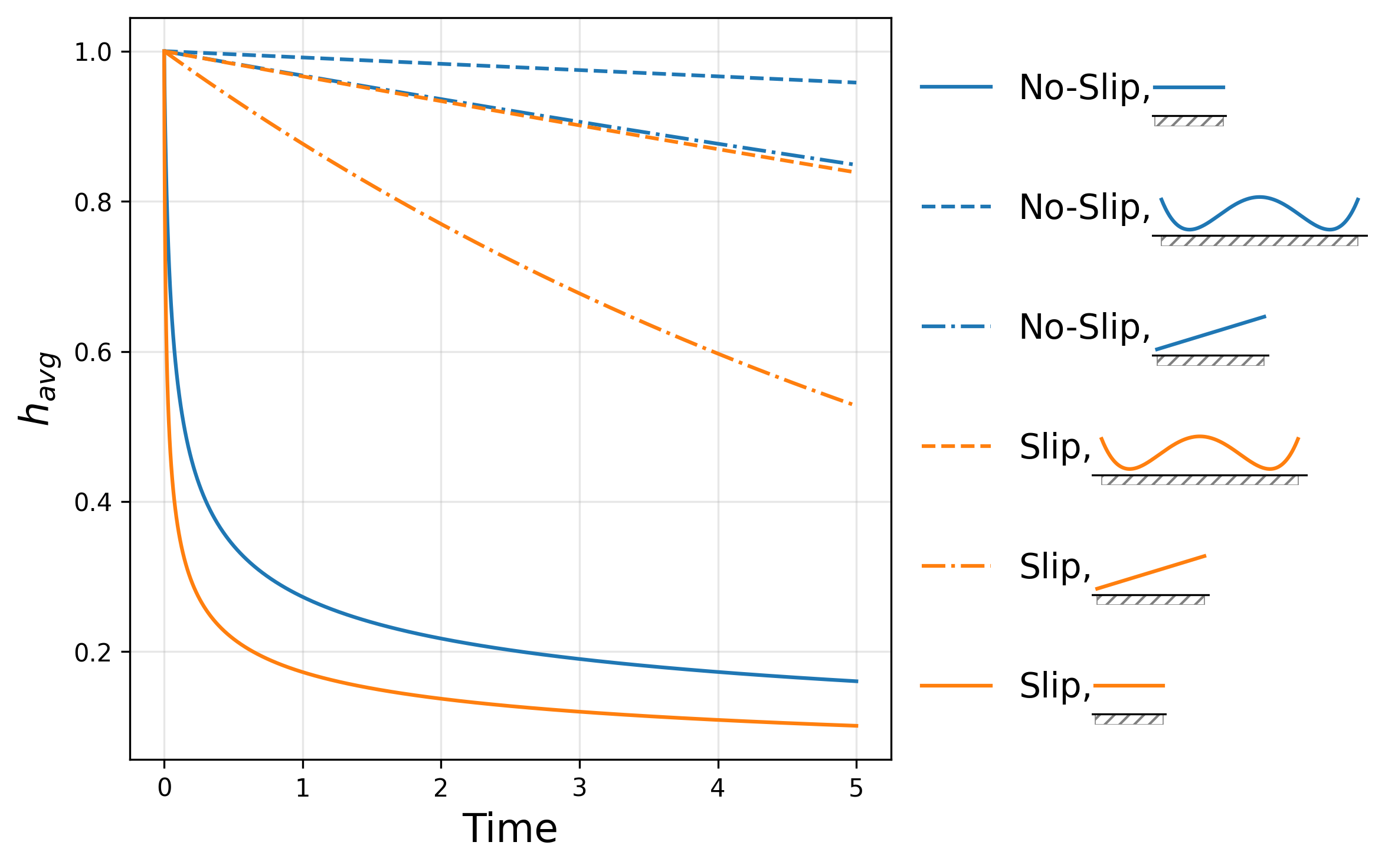}
    \caption{Variation in drainage of films of different morphologies when drained by the same force. Time is scaled by $\tau$, and average film thickness, $h_{avg}$ is scaled by the initial film thickness, $h_0$.}
    \label{fig:drainagetime}
\end{figure}

Scaling, $h$ with initial average film $h_0$, radial coordinate, $r$, with $R_c$ and time, $t$, with $\tau = \frac{24 \pi \mu R_c^4}{F  h_0^3}$, we get:
\begin{equation}
\frac{dh_{max}}{dt} \int_0^1 r \int_r^1 \frac{1}{r h(r,t)^4} \int_0^r r f(r)drdrdr + (3\beta -2) = 0
\label{eq:scaled-dhmaxdt}
\end{equation}
Solving, Eq. \ref{eq:scaled-dhmaxdt} numerically, we obtain the variation in $h_{max}$ and the corresponding $h_{avg}$ with time. We compare the variation in the drainage dynamics for different geometries and boundary conditions using a simplistic thin-film thickness evolution equation for $h_{max}$, assuming the profile shape remains the same, i.e., flat, $f(r) = 1$, linear, $1-r$, or double belled, $(1-r^2)^2$, during the evolution. Figure \ref{fig:drainagetime} shows the evolution of the average film thickness, indicating the drainage dynamics. For a flat film, the no-slip drainage is slower compared to the corresponding evolution when slip boundary condition is imposed at the 
bubble-liquid interface. Ratio of time taken to arrive at $h_{avg} = 0.2$ for no-slip to slip boundary condition for flat film is, $\sim 2.5/0.8 = 3.5$.  However, for a double-welled profile $f(r) = (1-r^2)^2$, even for slip boundary conditions, drainage is significantly slower compared to the flat film case with no-slip boundary conditions.

Further, comparing the evolution of double-welled profile with no-slip boundary condition, the drainage time is several order more compared to the corresponding evolution for slip boundary condition (ratio of time it takes to reach $h_{avg} = 0.9$ is $\sim 12.878/3.05 = 4.22$). The ratio of times to reach $h_{avg} = 0.9$ with no-slip conditions for flat film to double-welled film is $\sim 12.878/0.004 = 3.2\times 10^3$, suggesting a significant increase in drainage time for different geometries with the same boundary conditions.

\bibliographystyle{elsarticle-num} 
\bibliography{bibfile}

\end{document}